\documentclass[12pt]{article}
\usepackage{a4}
\usepackage{epstopdf}
\usepackage{amssymb}
\usepackage{bm}
\usepackage{texdraw}
\usepackage{graphicx}
\usepackage{appendix}
\def\gammab{{\mbox{\boldmath $\gamma$}}}

\def\comment#1{}
\def\y{\zeta}

\def\ev{v}

\def\ktheta{(\hbar k\theta)}
\def\blambda{\,\bar{}\!\!\lambda}

\title{On Electron-Positron Pair Production by a Spatially Nonuniform Electric Field}
\date{}
\author{A. Chervyakov
\footnote{alex{\_}cherv@yahoo.de~,
on leave from JINR, Dubna, Russia}
\,
and
H.~Kleinert
\footnote{h.k@fu-berlin.de~,
http://www.klnrt.de}
~\\
Institut f\"ur Theoretische Physik,\\
Freie Universit\"at Berlin,\\
Arnimallee 14, D-14195 Berlin}

\begin{document}
\maketitle

\begin{abstract}
A detailed analysis of electron-positron pair creation induced by a spatially nonuniform and static 
electric field from vacuum is presented. A typical example is provided by the Sauter potential.
For this potential, we derive the analytic expressions for vacuum decay and pair production rate 
accounted for the entire range of spatial variations. In the limit of a sharp step, we recover 
the divergent result due to the singular electric field at the origin. The limit of a constant 
field reproduces the classical result of Euler, Heisenberg and Schwinger, if the latter is properly 
averaged over the width of a spatial variation. The pair production by the Sauter potential is described for 
different physical regimes from weak to strong fields. In all these regimes, the locally constant-field 
rate is shown to be the upper limit. 
\end{abstract}

\section{Introduction}
Since Klein's formulation~\cite{klein} of his gedanken experiment in 1929 and its interpretation
by Sauter~\cite{sauter}, Heisenberg and Euler~\cite{heeu}, and Hund~\cite{hund} it became clear 
that the vacuum of quantum electrodynamics (QED) is unstable against the electron-positron pair 
production in the present of external electromagnetic field. The first discussions were all done 
at the level of first-quantized field theory. The results were rederived within second-quantized 
field theory in the one-loop approximation by Schwinger~\cite{schwinger} for a constant electromagnetic 
field, and Nikishov~\cite{nikishov1,nikishov2,nikishov3} and Narozhny and Nikishov~\cite{nar1,nar2} 
for more general field configurations. The two-loop radiative corrections were calculated by 
Ritus~\cite{ritus1,ritus2} and Lebedev and Ritus~\cite{ritus3}.
The spin-statistics connection in QED with unstable vacuum was discussed by Feynman~\cite{feynman}.
A general discussion, detailed calculations and further references on this subject can be found
in Refs.~\cite{damour,hansen,grib,zuber,greiner1,greiner2,gitman,kl,dunne1}.
For recent developments and the modern state of the problem, see Refs.~\cite{ruffini,dunne}.

In a constant electric field $E$, the probability for vacuum decay via pair creation in unit volume per unit time
(i.e., the vacuum decay rate) can be written as
\begin{eqnarray}
w^{\rm cf} (E) = \frac{c}{4\pi^{3}\,\blambda_{e}^{4}}\left(\frac{E}{E_{c}}\right)^{2}\sum^{\infty}_{n=1}
\frac{1}{n^2}\exp\left(- n\pi\frac{E_{c}}{E}\right)\,,
\label{0.1}\end{eqnarray}
where $E_{c}\equiv m^2 c^3/|e|\hbar\simeq 1.3\times 10^{18}\,{\rm V}/{\rm m}$ is the critical field,
whose work to move an electron over the Compton wavelength $\blambda_{e}=\hbar /m c$ being equal 
to its rest energy $mc^2$ and $\blambda_{e}^{4}/c\simeq 7.3\times 10^{-59}\,{\rm m}^{3}{\rm s}$
is the Compton space-time volume. The rate~(\ref{0.1}) results from the imaginary part of the one-loop  
Euler-Heisenberg Lagrangian~\cite{heeu}. The formula~(\ref{0.1}) was anticipated by Sauter~\cite{sauter} and 
derived by Euler and Heisenberg~\cite{heeu} and, more elegantly, by Schwinger~\cite{schwinger}.
It is presently often referred to as Schwinger formula, and the pair creation process as Schwinger mechanism 
of pair production. The process can be understood within the Dirac picture of vacuum as a quantum tunneling of 
electron through an electrostatic potential barrier created by the electric field. Its rate is, however, 
exponentially suppressed for presently accessible field strengths $E$ achieved roughly $4\times 10^{14}\,{\rm V}/{\rm m}$ 
by optical lasers~\cite{ring}. This leads to the present impossibility of observing
the pair creation for which the extraordinary strong electric field strengths $E$ of the order or above the 
critical value $E_{c}\simeq 1.3\times 10^{18}\,{\rm V}/{\rm m}$ are required. 

While creating such strong macroscopic fields in the laboratory is impossible today, it
will hopefully be achieved in the near future after further advances in the laser technology.
Indeed, the energy extraction and the optical focusing can be improved considerably in the
X-ray free electron lasers to produce the electric field of the order of $E\sim 10^{-2}\,E_{c}$
capable of yielding a sizeable effect of pair production~\cite{hei,bul1}.
Under these circumstances, we could get a significant progress in understanding the properties
of QED beyond the scope of perturbation theory in the strong-field regime.
Of course, such strong macroscopic fields in the laboratory
would never be as uniform as those originally considered in the Schwinger mechanism of pair creation.
Their spatial and temporal modulations must be taken into account
in optical and, especially, in X-ray free electron lasers (for review, see Ref.~\cite{ring}
and for recent estimates of pair production created by laser fields, Refs.~\cite{popov4,bul2,bul3,gitman2,dunne4,dunne5}).
Apart from these, enormous electromagnetic and gravitational
fields are expected in the powerful Gamma Ray Bursts, strong Coulomb fields in the
relativistic heavy-ion collisions and nonabelian gauge fields in QCD, for which
the Schwinger mechanism of pair creation is also relevant~\cite{ruffini}.

Going beyond the constant fields in space and time is therefore an important goal in studying pair
creation. First estimates of the impact of spatial and temporal inhomogeneities on pair production 
can be found in Refs.~\cite{nikishov1,nikishov2} for spatially varying fields, 
in Refs.~\cite{nar1,nar2,brezin,popov1,popov2,popov3,most} for oscillating electric fields, and 
in Ref.~\cite{wang} for an electric field enclosed by conducting plates. 
Recent studies of pair production by inhomogeneous electromagnetic fields including developments of many 
useful approximate schemes such as WKB method, world-line formalism, instanton and Monte Carlo techniques
can be found in Refs.~\cite{gitman1,dunne2,popov5,kim1,schub1,gies1,schub2,kim2,kim2a,kl1,kim3,kim4,cherv,gies2}.
The back-reaction of produced pairs on the external field has been also taken into account in 
Refs.~\cite{cooper1,cooper2}, and more recently in Ref.~\cite{akhmedov}.

For an arbitrary shape of background field the study of pair creation becomes rather involved, 
so that only special field configurations have been considered. In fact, one has to find 
the scattering and pair production probability densities in terms of solutions of the 
single-particle Klein-Gordon or Dirac equation in the presence of external field yielding
the Bogoliubov transformation coefficients within the $S$-matrix formalism. In addition, 
the energy-momentum integral over the level-crossing (Klein) region involving logarithm of the reflection 
coefficient must be performed. In this way, one can connect in a most transparent and efficient way the 
one-particle Dirac theory, in which the Klein paradox appeared with the second-quantized field theory, 
in which it was resolved satisfactorily.

In this paper, we examine the problem of pair creation of Dirac particles in a spatially
nonuniform and static electric field ${\bf E} (z) = (0, 0, E(z))$ directed in the $z$-axis and 
varied only along this axis, where $E(z)$ is the Sauter field of the form
\begin{eqnarray}
E (z) = \frac{E_{0}}{\cosh^2 kz}\,,
\label{0.3}\end{eqnarray}
with the peak $E_{0}=vk/|e|$ at $z=0$. The parameters $v$ and $1/k$ are associated with
the height and the width of the electrostatic potential $V(z) = e\phi (z)$ created by the field
$E(z)=-\phi' (z)$, respectively. Explicitly, it reads
\begin{eqnarray}
V(z) = e\phi (z) = v\tanh kz\,,
\label{0.2}\end{eqnarray}
with $v, k > 0$. The limit $k\rightarrow\infty$ leads to a sharp step, whereas the limit
$k\rightarrow 0$ with $v k$ fixed reproduces the linear potential caused by a constant field $E_0$.

With no magnetic field, we use the gauge $A^{\mu} (t,{\bf x_\perp},z) = (\phi (z), 0, 0,0)$ in which
the Dirac equation allows the full separation of variables. As a result, the problem of pair production 
is reduced to an exactly solvable quantum-mechanical problem~\cite{nikishov1,nikishov2,nikishov3} 
of the scattering of a Dirac particle on the static potential barrier~(\ref{0.2}). The problem is recapitulated 
in Section~2. This provides us with exact expressions for reflection and transmission coefficients in terms 
of scattering data for both the sub- and the supercritical values of the potential~(\ref{0.2}). For supercritical 
barriers with the level crossing between positive and negative energy levels, the Klein paradox is connected 
with the vacuum instability via the electron-positron pair production. In Section~3, we discuss how the reflection 
and transmission amplitudes are related to the pair production rates, first for a certain particular state, 
and then by taking all these states into account. In particular, the total vacuum decay rate due to 
pair creation is expressed as an energy and transverse momentum integral over the level-crossing (Klein) 
region involving the logarithm of reflection coefficient. 

The quantum-mechanical scattering amplitudes on the potential barrier~(\ref{0.2}) have been known 
since Suater's work~\cite{sauter}, and their relation to the pair production probabilities since 
Nikishov's work~\cite{nikishov1,nikishov2,nikishov3}. 
In order to find the pair production rates, we must however complete that work by evaluating the energy-momentum 
integral for a given local probabilities. The method of calculating such an integral has been developed recently in 
Ref.~\cite{cherv}. In Section~3, we apply this method to perform the rotationally invariant integral over transverse 
momenta in three dimensions exactly. The remaining integral over the energy, after an appropriate change 
of variable, provides us with simple spectral formulas for the vacuum decay rate and also for the pair production rate.
The analytic expressions for these rates are derived in Section~4 for the entire range of the parameters
$v$ and $k$ of the supercritical potential~(\ref{0.2}). This allows us to access different physical regimes
from weak to strong fields.

In the weak-field regime corresponding to presently available field strengths, the supercritical potential~(\ref{0.2}) 
can only create the pairs near the constant limit. The pair production rate is very small and is described 
by the semiclassical expressions of Refs.~\cite{schub1,gies1,schub2,kim2,kim2a,kl1} which are recovered and slightly 
corrected in this paper. The constant-field limit of the vacuum decay rate is found to be different from 
the Schwinger formula~(\ref{0.1}). It agrees, however, with formula~(\ref{0.1}), if the latter is properly
averaged over the width of a spatial variation. As a such, this limit represents the locally constant-field rate.
The explanation as well as the detailed comparison of the two constant-field rates can be found in Appendix~A. 

With increasing field strength from weak- to strong-field regime, the supercritical potential~(\ref{0.2})
creates the pairs much more intensively over entire field width extended from sharp to the constant limit.
The corresponding expression for the vacuum decay rate becomes very large but interpolates analytically 
between these limits always below the locally constant-field rate. The latter is therefore the upper limit for
the spatially nonuniform electric field~(\ref{0.3}).
 
\section{Barrier scattering and pair production}
Consider a relativistic spin-$1/2$ particle of charge $e$ ($e=-|e|$ for electron),
mass $m$ and energy $\varepsilon$ moving along the $z$-axis in the field~(\ref{0.3}) of the potential
barrier~(\ref{0.2}). The energy and momentum of a particle in this potential read ($c = 1$):
\begin{eqnarray}
p_{0} (z) =\varepsilon - V(z),\quad p_3 (z) = \sqrt{p^{2}_{0} (z) - (p^{2}_{\perp} + m^2)},\quad
p^{2}_{\perp} \equiv p^2_1 + p^2_2\,.
\label{1.1}\end{eqnarray}
In the transverse direction the particle propagates freely as a plane wave
$\exp[ i({\bf p_\perp} {\bf x_\perp} - \varepsilon \,t)/\hbar]$.
Its wave function can therefore be separated as a product of this plane wave
with a four-component spinor field $\psi (z)$ satisfying the one-dimensional Dirac equation
\begin{eqnarray}
\left[\gamma^{0} p_{0} (z) + \gammab^\perp {\bf p_\perp} + i\hbar\gamma^{3}\partial_{z}
- m\right]\psi (z) = 0\,,
\label{1.2}\end{eqnarray}
where $\gamma^{\mu}$ are the Dirac matrices. The solution of Eq.~(\ref{1.2}) is represented
in the following form
\begin{eqnarray}
\psi (z) = \left[\gamma^{0} p_{0} (z) + \gammab^\perp {\bf p_\perp}
+ i\hbar\gamma^{3}\partial_{z} + m\right]\chi (z)\,,
\label{1.3}\end{eqnarray}
where the four-component spinor field $\chi (z)$ satisfies the second-order Dirac equation
\begin{eqnarray}
\chi '' (z) + \frac{1}{\hbar^2}\left[p_{3}^{2} (z) - i\hbar e E(z)(\gamma^{0}\gamma^{3})\right]\chi (z) = 0\,.\label{1.4}
\end{eqnarray}
This has two sets of solutions, each associated with two spin states $\sigma =1,2$ of a particle
\begin{eqnarray}
\chi (z)\rightarrow\chi_{\sigma}(z) = \left\{
\begin{array}{ccc}
\varphi_{+1} (z)\,u_{\sigma}&\mbox{,}&\sigma = 1,2\\&\\
\varphi_{-1} (z)\,v_{\sigma}&\mbox{,}&\sigma = 1,2
\end{array}\right.\,,
\label{1.5}\end{eqnarray}
where the constant spinors $u_\sigma$ and $v_\sigma$ are eigenvectors of the matrix $(\gamma^{0}\gamma^{3})$
with eigenvalues $+1$ and $-1$, respectively. In the Dirac representation of $\gamma$-matrices, these read explicitly,
\begin{eqnarray}
u_{1} = \frac{1}{\sqrt{2}}\left(
\begin{array}{r}
0\\
1\\
0\\
-1\\
\end{array}\right),
\quad u_{2} = \frac{1}{\sqrt{2}}\left(
\begin{array}{c}
1\\
0\\
1\\
0\\
\end{array}\right),
\quad v_{1} = \frac{1}{\sqrt{2}}\left(
\begin{array}{r}
1\\
0\\
-1\\
0\\
\end{array}\right),
\quad v_{2} = \frac{1}{\sqrt{2}}\left(
\begin{array}{c}
0\\
1\\
0\\
1\\
\end{array}\right)\,,
\label{1.7}\end{eqnarray}
with the normalization
$u^{\dagger}_{\sigma}u_{\sigma'} = \delta_{\sigma\sigma'}$, $v^{\dagger}_{\sigma}v_{\sigma'} =
\delta_{\sigma\sigma'}$ and $u^{\dagger}_{\sigma}v_{\sigma'} = 0$, $v^{\dagger}_{\sigma}u_{\sigma'} = 0$.

The functions $\varphi_{\pm 1} (z)$ in Eq.~(\ref{1.5}) obey the Schr\"odinger-like equations
\begin{eqnarray}
\varphi_{s}'' (z) + \frac{1}{\hbar^2}\left[p_{3}^{2} (z) - i s\hbar e E(z)\right]\varphi_{s} (z) = 0\,,
\,\,s = \pm 1\,.
\label{1.6}\end{eqnarray}
For each $s$, there are two independent solutions which behave asymptotically as
plane waves outside the potential barrier~(\ref{0.2}).
Let us introduce the initial and final values of the particle energy far to the left
and to the right of the potential
\begin{eqnarray}
p_0^{(\mp)}\equiv\left.p_0 (z)\right|_{z\rightarrow\mp\infty} = \varepsilon\pm v\,,
\label{1.1a}\end{eqnarray}
and the corresponding momenta
\begin{eqnarray}
p_3^{(\mp)}\equiv\left.p_3 (z)\right|_{z\rightarrow\mp\infty} = \sqrt{p_0^{(\mp)}{}^2 - (p^{2}_\perp + m^2)}\,,
\label{1.1b}\end{eqnarray}
where we take the positive square roots. We assume also that momenta $p_3^{(\mp)}$ in Eq.~(\ref{1.1b})
are real thus restricting the asymptotic energies to $|p_0^{(\mp)}| > \sqrt{p^{2}_\perp + m^2}$.

For both signs $s=\pm 1$, the independent solutions possess the same asymptotic behavior
at $z\rightarrow -\infty$ as well as at $z\rightarrow +\infty$ due to
vanishing the electric field~(\ref{0.3}). Let ${}_{\pm}\varphi_{s}(z)$ be the two
independent solutions that describe the ingoing $(+)$ and the outgoing $(-)$ plane waves at
$z\rightarrow -\infty$,
\begin{eqnarray}
\left.{}_{\pm}\varphi_{s}(z)\right|_{z\rightarrow -\infty}
\approx  e^{\pm ip_3^{(-)}z/\hbar}\,,\quad s =\pm 1\,,
\label{1.11al}
\end{eqnarray}
whereas ${}^{\pm}\varphi_{s}(z)$ the two independent solutions that describe
the ingoing $(+)$ and the outgoing $(-)$ plane waves at $z\rightarrow +\infty$,
\begin{eqnarray}
\left.{}^{\pm}\varphi_{s}(z)\right|_{z\rightarrow +\infty}
\approx e^{\pm ip_3^{(+)}z/\hbar}\,,\quad  s =\pm 1\,.
\label{1.11br}\end{eqnarray}
Depending on two projections $r =\pm$ of momenta $p_3^{(\mp)}$ onto the $z$-axis,
we obtain two complete sets of solutions $\left\{{}_{r}\varphi_{s}\right\}$
and $\left\{{}^{r}\varphi_{s}\right\}$ for equations~(\ref{1.6}) satisfying
the asymptotic conditions~(\ref{1.11al}) and~(\ref{1.11br}), respectively.
By virtue of linearity of equations~(\ref{1.6}), each function from the one set
$\left\{{}_{\pm}\varphi_{s}\right\}$ can be expressed as a linear combination of the both functions
from the another set $\left\{{}^{\pm}\varphi_{s}\right\}$ and vice versa. The coefficients of these
combinations are then related to the reflection and transmission amplitudes of a particle in a scattering problem.
Explicitly, these can be found for potentials for which the equations~(\ref{1.6}) are solvable exactly.

For each $s$, the functions $\left\{{}_{\pm}\varphi_{s}\right\}$ as well as the functions
$\left\{{}^{\pm}\varphi_{s}\right\}$ are linearly independent. For each $r$, the functions
$\left\{{}_{r}\varphi_{\pm 1}\right\}$ as well as the functions $\left\{{}^{r}\varphi_{\pm 1}\right\}$
are related to each other. The relations for ${}^{r}\varphi_{+1}(z)$ and ${}^{r}\varphi_{-1}(z)$
follow from Eqs.~(\ref{1.6}) and~(\ref{1.11br}). Explicitly, these read
\begin{eqnarray}
\left[i\hbar\partial_{z} + p_{0} (z)\right]\,{}^{\pm}\varphi_{+1}(z) &=&\left[p_0^{(+)} \mp p_3^{(+)} \right]\,{}^{\pm}\varphi_{-1}(z)\,,\label{1.15a}\\
\left[i\hbar\partial_{z} - p_{0} (z)\right]\,{}^{\pm}\varphi_{-1}(z) &=& -\left[p_0^{(+)} \pm p_3^{(+)} \right]\,{}^{\pm}\varphi_{+1}(z)\,.
\label{1.15b}\end{eqnarray}
Similar relations can be found for ${}_{r}\varphi_{+1}(z)$ and ${}_{r}\varphi_{-1}(z)$
with the help of Eqs.~(\ref{1.6}) and~(\ref{1.11al}).

According to Eq.~(\ref{1.5}), each complete set of solutions $\left\{{}_{r}\varphi_{s}\right\}$
and $\left\{{}^{r}\varphi_{s}\right\}$ provides eight solutions for the second-order equation~(\ref{1.4})
with only four solutions being linearly independent. Using for instance
$\left\{{}^{r}\varphi_{s}\right\}$, we obtain eight solutions explicitly as
${}^{r}\varphi_{+1} (z)\,u_{\sigma}$ and ${}^{r}\varphi_{-1} (z)\,v_{\sigma}$ with $r=\pm$
and $\sigma = 1,2$, where the functions ${}^{r}\varphi_{+1} (z)$ and ${}^{r}\varphi_{-1} (z)$
obey Eqs.~(\ref{1.15a}) and~(\ref{1.15b}). Thus, we can use either ${}^{r}\varphi_{+1} (z)\,u_{\sigma}$,
or ${}^{r}\varphi_{-1} (z)\,v_{\sigma}$ in order to obtain four independent solutions for the Dirac equation~(\ref{1.2}) 
with the help of Eq.~(\ref{1.3}). Substituting for instance
${}^{\pm}\varphi_{-1} (z)\,v_{\sigma}$ into Eq.~(\ref{1.3}) and applying Eqs.~(\ref{1.15a})
and~(\ref{1.15b}) yields one complete set of solutions as
\begin{eqnarray}
{}^{\pm}\psi_{\sigma}(z) &=& \left[\gamma^{0} p_{0} (z) + \gammab^\perp {\bf p_\perp}
+ i\hbar\gamma^{3}\partial_{z} + m\right]\,{}^{\pm}\varphi_{-1} (z)\,v_{\sigma}\nonumber\\
&=& {}^{\pm} N\left\{{}^{\pm}\varphi_{-1}(z)\,\xi_{\sigma} +
\left[p_0^{(+)} \pm p_3^{(+)}\right]{}^{\pm}\varphi_{+1}(z)\,\eta_{\sigma}\right\}\,,\quad\sigma = 1,2\,,
\label{1.8}\end{eqnarray}
where $\xi_{\sigma}\equiv (\gamma^\perp p_{\perp} + m)v_{\sigma}$ and $\eta_{\sigma}\equiv (\gamma^{0}v_{\sigma})$
for $\sigma = 1,2$ and ${}^{\pm}N$ are normalization constants to be determined later.
Similarly, the another complete set of solutions for the Dirac equation~(\ref{1.2}) reads
\begin{eqnarray}
{}_{\pm}\psi_{\sigma}(z) &=& \left[\gamma^{0} p_{0} (z) + \gammab^\perp {\bf p_\perp} 
+ i\hbar\gamma^{3}\partial_{z} + m\right]\,{}_{\pm}\varphi_{-1} (z)\,v_{\sigma}\nonumber\\
&=& {}_{\pm} N\left\{{}_{\pm}\varphi_{-1}(z)\,\xi_{\sigma} +
\left[p_0^{(-)} \pm p_3^{(-)}\right]{}_{\pm}\varphi_{+1}(z)\,\eta_{\sigma}\right\}\,,\quad\sigma = 1,2\,,
\label{1.9}\end{eqnarray}
with the normalization constants ${}_{\pm}N$.
In Eqs.~(\ref{1.8}) and~(\ref{1.9}), the spinor fields with $\sigma\neq\sigma'$ are orthogonal $\psi^{\dagger}_{\sigma} (z)\psi_{\sigma'} (z)= 0$.

In order to solve equations~(\ref{1.6}) with the potential~(\ref{0.2}), we set
\begin{eqnarray}
p_3 ^{(-)} \equiv 2\hbar k\mu\,,\quad
p_3 ^{(+)}\equiv  2\hbar k\nu\,,\quad
\label{1.12}\end{eqnarray}
where
\begin{eqnarray}
\mu^2 - \nu^2 = \lambda\,\frac{\varepsilon}{\hbar k}\,,
\label{1.13}\end{eqnarray}
with
\begin{eqnarray}
\lambda\equiv \frac{v}{\hbar k} > 0\,.
\label{1.14}\end{eqnarray}
In these notations, equations~(\ref{1.6}) become
\begin{eqnarray}
\varphi_{\pm 1} '' (z) + \frac{k^2}{\cosh^2 k z}
\left[\nu^2\,e^{2k z} + \mu^2\,e^{-2k z} + \left(\nu^2 + \mu^2 -
\lambda^2\right) \pm i\lambda\right]\varphi_{\pm 1} (z) = 0\,.
\label{1.6a}\end{eqnarray}
Thus, the solution $\varphi_{+1} (z)\equiv \varphi (z,-\lambda)$
can be obtained from the solution $\varphi_{-1} (z)\equiv \varphi (z,\lambda)$
via the interchanging $\lambda\rightarrow -\lambda$ with $\lambda >0$.
The function $\varphi (z,\lambda)$ satisfies
\begin{eqnarray}
\varphi '' (z,\lambda) + \frac{k^2}{\cosh^2 k z}
\left[\nu^2\,e^{2k z} + \mu^2\,e^{-2k z} + \left(\nu^2 + \mu^2 -
\lambda^2\right) - i\lambda\right]\varphi (z,\lambda) = 0\,.
\label{1.15}\end{eqnarray}

Equation~(\ref{1.15}) is solvable exactly in terms of the hypergeometric function~\cite{nikishov2}.
For this equation, we are looking for two independent solutions ${}^{+}\varphi (z,\lambda)$
and ${}^{-}\varphi (z,\lambda)$ with fixed asymptotic behaviors at $z\rightarrow +\infty$:
\begin{eqnarray}
{}^{\pm}\varphi (z,\lambda)\approx \exp\left(\pm 2ik\nu z\right)\,,\quad z\rightarrow + \infty\,,
\label{1.16r}\end{eqnarray}
where the normalization can be neglected. Let us introduce the new dimensionless variable
\begin{eqnarray}
\y = - \exp(-2kz)\,,
\label{1.17}\end{eqnarray}
running from $-\infty$ to $0$. To find ${}^{+}\varphi (z,\lambda)$,
we use the substitution
\begin{eqnarray}
{}^{+}\varphi (\y,\lambda) = (-\y)^{-i\nu}\,f (\y,\lambda)\,,
\label{1.18}\end{eqnarray}
where the function $f(\y,\lambda)$ satisfies
\begin{eqnarray}
\y f''(\y,\lambda) + (1-2i\nu) f'(\y,\lambda) + \left[\frac{(\lambda^2 + i\lambda)}{(1-\y)^2}
- \frac{(\mu^2 - \nu^2)}{(1-\y)}\right]f (\y,\lambda) = 0 \,.
\label{1.19}\end{eqnarray}
To remove the singularity at $\y=1$ in Eq.~(\ref{1.19}), we replace
\begin{eqnarray}
f (\y,\lambda) = (1 - \y)^{i\lambda}\,w (\y,\lambda)\,.
\label{1.20}\end{eqnarray}
Eq.~(\ref{1.19}) becomes now a hypergeometric equation
\begin{eqnarray}
\!\!\y (1-\y) w''(\y,\lambda) + \left\{(1-2i\nu) - \left[(1-2i\nu) + 2i\lambda\right]\y\right\} w'(\y,\lambda)
+ \left[\left(\lambda - \nu\right)^2 - \mu^2 \right] w(\y,\lambda) = 0,
\label{1.21}\end{eqnarray}
where the function $w(\y,\lambda)$ tends to a constant for $\y\rightarrow 0\,(z\rightarrow +\infty)$.
With this condition the solution is the hypergeometric function
\begin{eqnarray}
w (\y,\lambda) = F \left[i\left(\lambda - \nu - \mu\right),\,
i\left(\lambda - \nu + \mu\right),\, 1 - 2i\nu\,;\,\y \,\right]\,
\label{1.22}\end{eqnarray}
up to some normalization factor.

The solution ${}^{+}\varphi (z,\lambda)$ of Eq.~(\ref{1.15}) reads finally,
\begin{eqnarray}
{}^{+}\varphi (z,\lambda) &=& (-\y)^{-i\nu}\,(1 - \y)^{i\lambda}\,F \left[i\left(\lambda - \nu - \mu\right),\,i\left(\lambda - \nu + \mu\right),\, 1 - 2i\nu\,;\,\y \,\right]\,,\nonumber\\
\quad\quad \y&\equiv& - \exp(-2kz)\,.
\label{1.23}\end{eqnarray}
For $z\rightarrow +\infty \,(\y\rightarrow 0)$, the function ${}^{+}\varphi (z,\lambda)$ satisfies the asymptotic condition in Eq.~(\ref{1.16r}).
To find its the asymptotic behavior for $z\rightarrow -\infty \,(\y\rightarrow -\infty)$, we use the Kummer transformation of
the hypergeometric function. This yields the superposition
\begin{eqnarray}
{}^{+}\varphi (z,\lambda) = a(\lambda)\,{}_{+}\varphi (z,\lambda)
+ b(\lambda)\,{}_{-}\varphi (z,\lambda)\,,
\label{1.24}\end{eqnarray}
with the amplitudes
\begin{eqnarray}
a(\lambda)&=&\frac{\Gamma (1-2i\nu)\,\Gamma (-2i\mu)}{\Gamma [i(\lambda -\nu -\mu)]\,\Gamma [1-i(\lambda + \nu +\mu)]}\,,\label{1.25}\\
b(\lambda)&=&\frac{\Gamma (1-2i\nu)\,\Gamma (2i\mu)}{\Gamma [i(\lambda - \nu + \mu)]\,\Gamma [1-i(\lambda + \nu - \mu)]}\,.
\label{1.26}\end{eqnarray}
Here the functions ${}_{+}\varphi (z,\lambda)$ and ${}_{-}\varphi (z,\lambda)$ are
two other independent solutions of Eq.~(\ref{1.15}) with fixed asymptotic behaviors at $z\rightarrow -\infty$:
\begin{eqnarray}
{}_{\pm}\varphi (z,\lambda)\approx \exp\left(\pm 2ik\mu z\right)\,,\quad z\rightarrow - \infty\,.
\label{1.27l}\end{eqnarray}
Their explicit forms will not be necessary further. To get the asymptotic behavior of
the function ${}^{+}\varphi (z,\lambda)$ at $z\rightarrow -\infty$ from Eq.~(\ref{1.24}),
we apply the conditions~(\ref{1.27l}) and obtain
\begin{eqnarray}
{}^{+}\varphi (z,\lambda)\approx a(\lambda)\exp\left(+2ik\mu z\right)
+ b(\lambda)\exp\left(-2ik\mu z\right)\,,\quad z\rightarrow - \infty\,.
\label{1.16al}\end{eqnarray}

In order to find the second solution ${}^{-}\varphi (z,\lambda)$, we use the invariance of Eq.~(\ref{1.15})
under the transformation involving a combination of complex conjugation with the substitution
$\lambda\rightarrow -\lambda$ for $\lambda >0$. Applying this transformation to the solution in
Eq.~(\ref{1.23}) yields
\begin{eqnarray}
{}^{-}\varphi (z,\lambda) &=& (-\y)^{i\nu}\,(1 - \y)^{i\lambda}\,F \left[i\left(\lambda + \nu + \mu\right),\,i\left(\lambda
+ \nu - \mu\right),\, 1 + 2i\nu\,;\,\y \,\right]\,,\nonumber\\
\quad\quad \y&\equiv& - \exp(-2kz)\,.
\label{1.28}\end{eqnarray}
For $z\rightarrow +\infty$, the solution ${}^{-}\varphi (z,\lambda)$ satisfies the asymptotic
condition in Eq.~(\ref{1.16r}). For $z\rightarrow -\infty$, we use the Kummer transformation
of the hypergeometric function in Eq.~(\ref{1.28}):
\begin{eqnarray}
{}^{-}\varphi (z,\lambda) = b^{\ast}(-\lambda)\,{}_{+}\varphi (z,\lambda)
+ a^{\ast}(-\lambda)\,{}_{-}\varphi (z,\lambda)\,,
\label{1.29}\end{eqnarray}
and the asymptotic conditions~(\ref{1.27l}). This yields
\begin{eqnarray}
{}^{-}\varphi (z,\lambda)\approx b^{\ast}(-\lambda)\exp\left(+2ik\mu z\right)
+ a^{\ast}(-\lambda)\exp\left(-2ik\mu z\right)\,,\quad z\rightarrow - \infty\,.
\label{1.16bl}\end{eqnarray}

The equation~(\ref{1.6a}) for $\varphi_{+1} (z)\equiv \varphi (z,-\lambda)$ can be obtained
from Eq.~(\ref{1.15}) via the interchanging $\lambda\rightarrow -\lambda$ for $\lambda >0$.
The corresponding solutions ${}^{+}\varphi (z,-\lambda)$ and ${}^{-}\varphi (z,-\lambda)$
are given by Eqs.~(\ref{1.23}) and~(\ref{1.28}) respectively, with the same interchanging.
For $z\rightarrow +\infty$, their asymptotic behaviors are coincident with the asymptotic behaviors
of the functions ${}^{\pm}\varphi (z,\lambda)$ in Eq.~(\ref{1.16r}):
\begin{eqnarray}
{}^{\pm}\varphi (z,-\lambda)\approx \exp\left(\pm 2ik\nu z\right)\,,\quad z\rightarrow + \infty\,.
\label{1.30r}\end{eqnarray}
Note that both equations~(\ref{1.6a}) take the same independent of $\lambda$ form in this limit.
Accordingly, the asymptotic behaviors of the second set of independent solutions
${}_{+}\varphi (z,-\lambda)$ and ${}_{-}\varphi (z,-\lambda)$ coincide with that for
the functions ${}_{\pm}\varphi (z,\lambda)$ in Eq.~(\ref{1.27l}):
\begin{eqnarray}
{}_{\pm}\varphi (z,-\lambda)\approx\exp\left(\pm 2ik\mu z\right)\,,\quad z\rightarrow - \infty\,.
\label{1.31l}\end{eqnarray}
We use these conditions to find the asymptotic behavior of the functions
${}^{\pm}\varphi (z,-\lambda)$ at $z\rightarrow -\infty$ from  Eqs.~(\ref{1.24}) and~(\ref{1.29})
with the replacement $\lambda\rightarrow -\lambda$. This yields
\begin{eqnarray}
&&{}^{+}\varphi (z,-\lambda)\approx a(-\lambda)\exp\left(+2ik\mu z\right)
+ b(-\lambda)\exp\left(-2ik\mu z\right)\,,\quad\!\! z\rightarrow -\infty\,,
\label{1.30al}\\
&&{}^{-}\varphi (z,-\lambda)\approx b^{\ast}(\lambda)\exp\left(+2ik\mu z\right)\,
+ \,a^{\ast}(\lambda)\exp\left(-2ik\mu z\right)\,,\quad\, z\rightarrow -\infty\,.
\label{1.30bl}\end{eqnarray}
In Eqs.~(\ref{1.30al}),~(\ref{1.30bl}) and~(\ref{1.29}),~(\ref{1.16bl}),
the amplitudes can be found from Eqs.~(\ref{1.25}) and~(\ref{1.26})
with the help of relations
\begin{eqnarray}
a (-\lambda)= \frac{(\mu +\nu +\lambda)}{(\mu +\nu -\lambda)}\,a (\lambda)\,,\quad
b (-\lambda) = \frac{(\mu -\nu -\lambda)}{(\mu -\nu +\lambda)}\,b (\lambda)\,,
\label{1.256}\end{eqnarray}
and their complex conjugated.

Having obtained the functions ${}^{\pm}\varphi (z,\lambda)$ and ${}^{\pm}\varphi (z,-\lambda)$,
we find from Eq.~(\ref{1.8}) four independent solutions of the Dirac equation~(\ref{1.2})
explicitly as
\begin{eqnarray}
{}^{\pm}\psi_{\sigma}(z) = \frac{{}^{\pm}\varphi (z,\lambda)\,\xi_{\sigma} +
\left[p_0^{(+)} \pm p_3^{(+)}\right]{}^{\pm}\varphi (z,-\lambda)\,\eta_{\sigma}
}{\sqrt{2p_3^{(+)}|p_0^{(+)} \pm p_3^{(+)}|}}\,,\quad\sigma = 1,2\,.
\label{1.32}\end{eqnarray}
Their asymptotic behaviors are fixed at $z\rightarrow +\infty$, where according to Eqs.~(\ref{1.16r})
and~(\ref{1.30r}) we obtain
\begin{eqnarray}
{}^{\pm}\psi_{\sigma}(z)\approx {}^{\pm}\psi^{R}_{\sigma}(z)\equiv\frac{\exp(\pm ip_3^{(+)}z/\hbar)
}{\sqrt{2p_3^{(+)}|p_0^{(+)} \pm p_3^{(+)}|}}\left[\xi_{\sigma} +
(p_0^{(+)} \pm p_3^{(+)})\eta_{\sigma}\right]\,,\quad
z\rightarrow +\infty\,,\quad\sigma = 1,2\,.
\label{1.33r}\end{eqnarray}
In turn, the asymptotic behaviors of four other independent solutions ${}_{\pm}\psi_{\sigma}(z)$
are fixed at $z\rightarrow -\infty$, where it follows
from Eq.~(\ref{1.9}) with the help of Eqs.~(\ref{1.27l}) and~(\ref{1.31l})
that
\begin{eqnarray}
{}_{\pm}\psi_{\sigma}(z)\approx {}_{\pm}\psi^{L}_{\sigma}(z)\equiv\frac{\exp(\pm ip_3^{(-)}z/\hbar)
}{\sqrt{2p_3^{(-)}|p_0^{(-)} \pm p_3^{(-)}|}}\left[\xi_{\sigma} +
(p_0^{(-)} \pm p_3^{(-)})\eta_{\sigma}\right]
\,,\quad
z\rightarrow -\infty\,,\quad\sigma = 1,2\,.
\label{1.34l}\end{eqnarray}
The wave functions in Eq.~(\ref{1.32}) are normalized in such a way
that the absolute value of current density along the $z$-axis evaluated with respect to
these functions for the states with given $\sigma$, $\varepsilon$ and ${\bf p}_{\perp}$ is equal to unity.
By virtue of Eq.~(\ref{1.2}), this quantity is the $z$-independent and can therefore be evaluated with
the help of the asymptotic expressions~(\ref{1.33r}) and~(\ref{1.34l}). The normalization is thus $|{}^{\pm}{\bar\psi}^{R}_{\sigma}(z)\gamma^{3}{}^{\pm}\psi^{R}_{\sigma}(z)|=
|{}_{\pm}{\bar\psi}^{L}_{\sigma}(z)\gamma^{3}\!\!{}_{\pm}\psi^{L}_{\sigma}(z)|=1$.

By means of asymptotic expressions~(\ref{1.34l}),
the asymptotic behaviors of the solutions ${}^{\pm}\psi_{\sigma}(z)$ in Eq.~(\ref{1.32})
at $z\rightarrow -\infty$ can be found from Eqs.~(\ref{1.16al}),~(\ref{1.30al}) and
Eqs.~(\ref{1.16bl}),~(\ref{1.30bl}), respectively. For ${}^{+}\psi_{\sigma}(z)$,
this yields
\begin{eqnarray}
{}^{+}\psi_{\sigma}(z)\approx I(\lambda)\,{}_{+}\psi^{L}_{\sigma}(z) +
R(\lambda)\,{}_{-}\psi^{L}_{\sigma}(z)\,,\quad z\rightarrow -\infty\,,\quad\sigma = 1,2\,,
\label{1.33al}\end{eqnarray}
where the amplitudes $I(\lambda)$ and $R(\lambda)$ are determined
independently of the spin state $\sigma =1,2$ with the help of relations~(\ref{1.256}) as
\begin{eqnarray}
I(\lambda)=\left(\frac{\mu\,|\mu +\nu +\lambda|}{\nu\,|\mu +\nu -\lambda|}\right)^{1/2} a(\lambda)\,,
\quad R(\lambda)=\left(\frac{\mu\,|\mu -\nu -\lambda|}{\nu\,|\mu -\nu +\lambda|}\right)^{1/2}b(\lambda)\,.
\label{1.35}\end{eqnarray}

Thus, we find the solution ${}^{+}\psi_{\sigma}(z)$ satisfying the asymptotic boundary conditions
\begin{eqnarray}
{}^{+}\psi_{\sigma}(z)\longrightarrow\left\{
\begin{array}{ccc}
I(\lambda)\,{}_{+}\psi^{L}_{\sigma}(z) +
R(\lambda)\,{}_{-}\psi^{L}_{\sigma}(z)&\mbox{,}&z\rightarrow -\infty\\&\\
{}^{+}\psi^{R}_{\sigma}(z)&\mbox{,}&z\rightarrow +\infty
\end{array}\right.\,.\label{1.36}
\end{eqnarray}
The process in Eq.~(\ref{1.36}) involves the electron wave of the energy $\varepsilon >0$
impacting from $z=-\infty$ with the amplitude $I(\lambda)$ on the potential
barrier~(\ref{0.2}) with $v>0$ at early times, which is partly reflected back to $z=-\infty$
with the amplitude $R (\lambda)$ at late times. The energy of electron far to the left
is therefore positive $p_0^{(-)}= (\varepsilon + v)\geq\sqrt{p^{2}_\perp + m^2}\geq 0$,
and its momentum $p_3 ^{(-)}$ is real and positive in this region.

In order to describe the plane wave solution involved in this process at $z=+\infty$,
we have to specify the height $v$ of the potential barrier~(\ref{0.2}).
For weak (subcritical) potentials with $v\leq \varepsilon -\sqrt{p^{2}_\perp + m^2}$, this plane wave
solution with the unit amplitude corresponds to a particle transmitted through the
potential barrier at late times and moving towards $z=+\infty$ with the positive energy
$p_0^{(+)}= (\varepsilon - v)\geq\sqrt{p^{2}_\perp + m^2}\geq 0$ and with the real and positive
momentum $p_3 ^{(+)}$.

However, for very strong (supercritical) potentials with the height
$v\geq \varepsilon +\sqrt{p^{2}_\perp + m^2}$ there are no longer particles in
the asymptotic region of the large positive $z$. In this case, the energy $\varepsilon$ of impinging
electron lies in the interval $-v +\sqrt{p^{2}_\perp + m^2}\leq\varepsilon\leq v-\sqrt{p^{2}_\perp + m^2}$
henceforth called the Klein region. The energy far to the right becomes now negative
$p_0^{(+)}=(\varepsilon - v)\leq -\sqrt{p^{2}_\perp + m^2}\leq 0$, while the momentum $p_3 ^{(+)}$
is real and, by our assumption, is still positive. For $v > \sqrt{p^{2}_\perp + m^2} > m$,
the positive energy levels far to the left overlaps the negative energy
levels far to the right in the Klein region. The plane wave solution ${}^{+}\psi^{R}_{\sigma}(z)$
with negative energy in Eq.~(\ref{1.36}) represents the transmitted wave via a quantum tunneling.
This right-moving wave propagates with the unit amplitude backwards in time and corresponds
to an incoming antiparticle coming in from $z=+\infty$ and moving forwards in time
towards the potential barrier. In the process~(\ref{1.36}), the incoming from the left at $z=-\infty$
particle annihilates the incoming from the right at $z=+\infty$ antiparticle.

With wave functions~(\ref{1.34l}), we obtain the incident and reflected electron currents
in the asymptotic region $z<0$ for the process~(\ref{1.36}) as follows
\begin{eqnarray}
{\bf j}^{\,{\rm p,\,in}}_{\,z<0}&\equiv&|I(\lambda)|^2\left({}_{+}{\bar\psi}^{L}_{\sigma}\gammab {}_{+}\psi^{L}_{\sigma}\right)
=|I(\lambda)|^2\left({}_{+}{\bar\psi}^{L}_{\sigma}\gamma^{3}\!\!{}_{+}\psi^{L}_{\sigma}\right)
{\bf\hat z}
= |I(\lambda)|^2\,{\bf\hat z}\,,\label{1.37i}\\
{\bf j}^{\,{\rm p,\,refl}}_{\,z<0}&\equiv&|R(\lambda)|^2\left({}_{-}{\bar\psi}^{L}_{\sigma}\gammab {}_{-}\psi^{L}_{\sigma}\right)
=|R(\lambda)|^2\left({}_{-}{\bar\psi}^{L}_{\sigma}\gamma^{3}\!\!{}_{-}\psi^{L}_{\sigma}\right)
{\bf\hat z}
= - |R(\lambda)|^2\,{\bf\hat z}\,,
\label{1.37r}\end{eqnarray}
where the superscript ${\rm p}$ denotes the particle state and, according to
Eqs.~(\ref{1.35}),~(\ref{1.25}) and~(\ref{1.26}),
\begin{eqnarray}
|I (\lambda)|^2 &=&\mp\frac{\sinh\pi(\lambda -\nu -\mu)\,\sinh\pi(\lambda +\nu +\mu)}
{\sinh2\pi\mu\,\sinh2\pi\nu}\geq 0\,,\label{1.38i}\\
|R (\lambda)|^2 &=&\mp\frac{\sinh\pi(\lambda -\nu +\mu)\,\sinh\pi(\lambda +\nu -\mu)}
{\sinh2\pi\mu\,\sinh2\pi\nu}\geq 0\,.
\label{1.38r}\end{eqnarray}
Here and below the upper/lower sign refers to (sub/super)critical potential barrier~(\ref{0.2}),
respectively. Note that $(\lambda\pm\nu -\mu)^{<}_{>}0$ for (sub/super)critical potential,
while $(\lambda\pm\nu +\mu)>0$ for the both potentials.

The current in the asymptotic region $z>0$ is evaluated with respect to wave functions~(\ref{1.33r}).
This yields
\begin{eqnarray}
{\bf j}^{\,{\rm p,\,tr/n,\,in}}_{\,z>0}\equiv\left({}^{+}{\bar\psi}^{R}_{\sigma}\gammab {}^{+}\psi^{R}_{\sigma}\right)
=\left({}^{+}{\bar\psi}^{R}_{\sigma}\gamma^{3}{}^{+}\psi^{R}_{\sigma}\right)
{\bf\hat z}
= \pm {\bf\hat z}\,,
\label{1.37t}\end{eqnarray}
where ${\bf j}^{\,{\rm p,\,tr}}_{\,z>0}$ is the transmitted towards $z=+\infty$
electron current and ${\bf j}^{\,{\rm n,\,in}}_{\,z>0}$ is the incident from $z=+\infty$
positron current directed in the positive/negative $z$-direction within the asymptotic region $z>0$ of
the (sub/super)critical potential barrier~(\ref{0.2}), respectively. The superscript ${\rm n}$
denotes the antiparticle state. According to Eqs.~(\ref{1.37i}),~(\ref{1.37r}) and~(\ref{1.37t}),
the conservation of current along the $z$-axis reads
\begin{eqnarray}
|I (\lambda)|^2 - |R (\lambda)|^2  = \pm1\,,
\label{1.39}\end{eqnarray}
with $|I (\lambda)|^2$ and $|R (\lambda)|^2$ are given in Eqs.~(\ref{1.38i}) and~(\ref{1.38r}).

The reflection coefficient is then defined by the ratio
\begin{eqnarray}
r\equiv\frac{|j^{\,{\rm p,\,refl}}_{\,z<0}|}{|j^{\,{\rm p,\,in}}_{\,z<0}|}
=\frac{|R (\lambda)|^2}{|I (\lambda)|^2} =
\frac{\sinh\pi(\lambda -\nu +\mu)\,\sinh\pi(\lambda +\nu -\mu)}
{\sinh\pi(\lambda -\nu -\mu)\,\sinh\pi(\lambda +\nu +\mu)}\geq0\,.
\label{1.40}\end{eqnarray}
It is the relative probability for elastic scattering of electron by (sub/super)critical potential barrier~(\ref{0.2}).
The transmission coefficient reads
\begin{eqnarray}
t\equiv\frac{|j^{\,{\rm p,\,tr/n,\,in}}_{\,z>0}|}{|j^{\,{\rm p,\,in}}_{\,z<0}|}
=\frac{1}{|I (\lambda)|^2} =
\mp\frac{\sinh2\pi\mu\,\sinh2\pi\nu}
{\sinh\pi(\lambda -\nu -\mu)\,\sinh\pi(\lambda +\nu +\mu)}\geq0\,.
\label{1.41}\end{eqnarray}
For supercritical potentials, the probability~(\ref{1.41})
is a relative probability for incoming electron to annihilate the incoming positron.
The coefficients~(\ref{1.40}) and~(\ref{1.41}) were first obtained by Sauter~\cite{sauter}.

From Eqs.~(\ref{1.39}),~(\ref{1.40}) and~(\ref{1.41}) we find the relation between reflection and
transmission coefficients for (sub/super)critical potential
\begin{eqnarray}
r \pm t =1\,.
\label{1.42}\end{eqnarray}
For supercritical potentials, the reflected electron current is therefore larger
than the incoming electron current, and even without the incoming electron, the
reflected current is exactly equal to the transmitted current. This is known as the
Klein paradox~\cite{klein} for which the single-particle description becomes inadequate,
since the time evolution operator of a single-electron wave function is no longer unitary
as far as $r\geq1$. Contrary,
within the second quantization the scattering matrix operator is always unitary and the
Klein paradox implies that the negative energy continuum of the Dirac vacuum
contributes to the both currents via production of electron-positron pairs under the influence
of supercritical potential. The probability~(\ref{1.41}) is then the relative probability for spontaneous
production of a single electron-positron pair.

To show this, we go back to our discussion above Eq.~(\ref{1.33al}) in order to find
the asymptotic behavior of the second independent solution ${}^{-}\psi_{\sigma}(z)$ in Eq.~(\ref{1.32})
at $z\rightarrow -\infty$ in terms of asymptotic expressions~(\ref{1.34l}). From now on,
we shall only consider the  supercritical potentials. In this case,
applying Eqs.~(\ref{1.16bl}),~(\ref{1.30bl}) to Eq.~(\ref{1.32}), we obtain
\begin{eqnarray}
{}^{-}\psi_{\sigma}(z)\approx - R^{\ast}(\lambda)\,{}_{+}\psi^{L}_{\sigma}(z) -
I^{\ast}(\lambda)\,{}_{-}\psi^{L}_{\sigma}(z)\,,\quad z\rightarrow -\infty\,,\quad\sigma = 1,2\,,
\label{1.33bl}\end{eqnarray}
where the amplitudes $I^{\ast}(\lambda)$ and $R^{\ast}(\lambda)$
are complex conjugated of the relations~(\ref{1.35}).

Thus, the solution ${}^{-}\psi_{\sigma}(z)$ has the following asymptotic behaviors at both infinities
\begin{eqnarray}
{}^{-}\psi_{\sigma}(z)\longrightarrow\left\{
\begin{array}{ccc}
-R^{\ast}(\lambda)\,{}_{+}\psi^{L}_{\sigma}(z) -
I^{\ast}(\lambda)\,{}_{-}\psi^{L}_{\sigma}(z)&\mbox{,}&z\rightarrow -\infty\\&\\
{}^{-}\psi^{R}_{\sigma}(z)&\mbox{,}&z\rightarrow +\infty
\end{array}\right.\,.
\label{1.44}\end{eqnarray}
The process~(\ref{1.44}) describes directly the steady state of the electron-positron
pair production by means of amplitudes of the previous process~(\ref{1.36}):
within the asymptotic region $z<0$ of the supercritical potential barrier~(\ref{0.2}) there are
the ingoing electron wave ${}_{+}\psi^{L}_{\sigma}(z)$ with the amplitude $-R^{\ast}(\lambda)$
and the outgoing electron wave ${}_{-}\psi^{L}_{\sigma}(z)$ with the amplitude $-I^{\ast}(\lambda)$.
This wave is accompanied by the outgoing positron wave ${}^{-}\psi^{R}_{\sigma}(z)$ with the unit amplitude
in the asymptotic region $z>0$. The corresponding currents far to the left are
\begin{eqnarray}
{\bf j}^{\,{\rm p,\,in}}_{\,z<0}&\equiv&|R(\lambda)|^2\left({}_{+}{\bar\psi}^{L}_{\sigma}\gammab {}_{+}\psi^{L}_{\sigma}\right)
=|R(\lambda)|^2\left({}_{+}{\bar\psi}^{L}_{\sigma}\gamma^{3}\!\!{}_{+}\psi^{L}_{\sigma}\right)
{\bf\hat z} = |R(\lambda)|^2\,{\bf\hat z}\,,\label{1.45i}\\
{\bf j}^{\,{\rm p,\,out}}_{\,z<0}&\equiv&|I(\lambda)|^2\left({}_{-}{\bar\psi}^{L}_{\sigma}\gammab {}_{-}\psi^{L}_{\sigma}\right)
=|I(\lambda)|^2\left({}_{-}{\bar\psi}^{L}_{\sigma}\gamma^{3}\!\!{}_{-}\psi^{L}_{\sigma}\right)
{\bf\hat z}
= - |I(\lambda)|^2\,{\bf\hat z}\,,
\label{1.45o}\end{eqnarray}
and, far to the right is
\begin{eqnarray}
{\bf j}^{\,{\rm n,\,out}}_{\,z>0}\equiv\left({}^{-}{\bar\psi}^{R}_{\sigma}\gammab {}^{-}\psi^{R}_{\sigma}\right)
=\left({}^{-}{\bar\psi}^{R}_{\sigma}\gamma^{3}{}^{-}\psi^{R}_{\sigma}\right)
{\bf\hat z} =  {\bf\hat z}\,,
\label{1.46o}\end{eqnarray}
with the current conservation being the same as in Eq.~(\ref{1.39}) for
supercritical potentials.

From Eqs.~(\ref{1.46o}) and~(\ref{1.45o}), the relative
probability for production of a single electron-positron pair is
\begin{eqnarray}
w\equiv\frac{|j^{\,{\rm n,\,out}}_{\,z>0}|}{|j^{\,{\rm p,\,out}}_{\,z<0}|}
=\frac{1}{|I (\lambda)|^2} =
\frac{\sinh2\pi\mu\,\sinh2\pi\nu}
{\sinh\pi(\lambda -\nu -\mu)\,\sinh\pi(\lambda +\nu +\mu)}\geq0\,,
\label{1.47}\end{eqnarray}
and therefore $w\equiv t$ with $t$ being the transmission coefficient for supercritical potentials
found before in Eq.~(\ref{1.41}).

The average rate of produced electron-positron pairs in the process~(\ref{1.44})
is then derived from Eqs.~(\ref{1.46o}) and~(\ref{1.45i}) as
\begin{eqnarray}
{\bar n}\equiv\frac{|j^{\,{\rm n,\,out}}_{\,z>0}|}{|j^{\,{\rm p,\,in}}_{\,z<0}|}
=\frac{1}{|R(\lambda)|^2} =
\frac{\sinh2\pi\mu\,\sinh2\pi\nu}
{\sinh\pi(\lambda -\nu +\mu)\,\sinh\pi(\lambda +\nu -\mu)}\geq0\,
\label{1.48}\end{eqnarray}
and, together with Eqs.~(\ref{1.40}) and~(\ref{1.41}), reads
\begin{eqnarray}
{\bar n} = t/r\,.
\label{1.49}\end{eqnarray}
It is the absolute probability for production of a single electron-positron pair in a given state
for which $t$ is the relative probability and $1/r$ is the probability that no pairs are created in this state.
This probability is called the vacuum-to-vacuum probability and is determined in such a way that
the total probability for production of zero and one pairs is unity $(1/r)(1 + t) = 1$ which follows from
Eq.~(\ref{1.42}) for supercritical potentials. Thus creating more than one pair is blocked by Eq.~(\ref{1.42})
in the agreement with the Pauli principle for fermions. Note, finally, that $1/r$ is the reflection coefficient
and $t/r$ is the transmission coefficient for the process~(\ref{1.44}).

\section{Vacuum decay and pair production rates}
In the previous section, the particle-antiparticle pair production has been described for each particular
state $n=(p, \sigma)$ which is completely specified by the total energy and transverse momentum
$p =\{\varepsilon,\,\bf{p_{\perp}}\}$, and also by a certain spin projection $\sigma$.
For these conserved quantum numbers, we shall treat all physical transitions related to each separate state $n$
as the independent events.

The exact solutions of a single-particle Dirac equation in external fields can be used
for calculating the scattering and pair production probabilities in quantum field theory~\cite{nikishov1,nikishov2,nikishov3}.
Consider the in- and out-going asymptotic solutions ${}_{\pm}\psi (z)$ for one-particle states given in Eq.~(\ref{1.34l})
as well as the in- and out-going asymptotic solutions ${}^{\pm}\psi (z)$ for one-antiparticle states given in Eq.~(\ref{1.33r}).
Their asymptotic forms are found after solving the Dirac equation~(\ref{1.2}) in the potential~(\ref{0.2}) exactly. As a result,
all these in- and out-going one-particle (antiparticle) states are related by Eqs.~(\ref{1.36}) and~(\ref{1.44}) with
the coefficients $I(\lambda)$ and $R(\lambda)$ and their complex conjugated given in Eq.~(\ref{1.35}), and also
in Eqs.~(\ref{1.25}) and~(\ref{1.26}). Accordingly, any field operator can be expanded either over all possible in-states or,
equivalently, over all possible out-states in terms of creation and annihilation operators. These are also separated as
in- and out-operators in such a way that the operator $a^{\rm in}_{n}/a^{\rm out}_{n}$ annihilates the in/out-going particle
with the wave function ${}_{+}\psi (z)/{}_{-}\psi (z)$, while the operator
$\left(b^{\rm in}_{n}\right)^{\dagger}/\left(b^{\rm out}_{n}\right)^{\dagger}$
creates the in/out-going antiparticle with the wave function ${}^{+}\psi (z)/{}^{-}\psi (z)$, respectively.
The key point~\cite{nikishov1,nikishov2,nikishov3}
is now that the asymptotic relations~(\ref{1.36}) and~(\ref{1.44}) between ${}_{+}\psi (z)/{}_{-}\psi (z)$ and ${}^{+}\psi (z)/{}^{-}\psi (z)$
determine completely the coefficients of the Bogoliubov transformation between the in- and out-going operators
\begin{eqnarray}
a^{\rm in}_{n} = -\frac{I(\lambda)}{R(\lambda)}\,a^{\rm out}_{n}-
\frac{1}{R(\lambda)}\,\left(b^{\rm out}_{n}\right)^{\dagger}\,,
\quad
\left(b^{\rm in}_{n}\right)^{\dagger} = -\frac{1}{R(\lambda)}\,a^{\rm out}_{n} - \frac{I^{\ast}(\lambda)}{R(\lambda)}\,
\left(b^{\rm out}_{n}\right)^{\dagger}\,,
\label{2.01}\end{eqnarray}
whose anticommutation relations preserve the current conservation~(\ref{1.39})
for supercritical potentials. Thus the Klein paradoxical result $|R (\lambda)|^{2}\geq 1$
of a single-particle treatment is a direct consequence of the fact that fermions obey the Pauli
principle and therefore quantized by anticommutators~\cite{hansen,grib}.

With these operators, one can construct two complete sets of in- and out-states from the two vacua.
These are the in-vacuum state $|0^{\rm in}_{n}\rangle$ and the out-vacuum state $|0^{\rm out}_{n}\rangle$
related for a certain $n$ by the unitary $S$-matrix operator $S_n$ as follows
$|0^{\rm in}_{n}\rangle = S_{n}|0^{\rm out}_{n}\rangle$.
The both are annihilated by the operators $a^{\rm in}_{n}$ and $a^{\rm out}_{n}$, respectively.
This allows one to calculate all necessary matrix elements.
For example, the absolute probability amplitude for pair creation in a given state is
\begin{eqnarray}
w^{\rm pair}_{n}\equiv\left\langle 0^{\rm out}_{n}\left|b^{\rm out}_{n}\,a^{\rm out}_{n}\right|0^{\rm in}_{n}\right\rangle =
-\frac{1}{I(\lambda)}\left\langle 0^{\rm out}_{n}\left|b^{\rm out}_{n}\,\left(b^{\rm out}_{n}\right)^{\dagger}
\right|0^{\rm in}_{n}\right\rangle =
-\frac{1}{I(\lambda)}\left\langle 0^{\rm out}_{n}\left|\,0^{\rm in}_{n}\right.\right\rangle\,,
\label{2.02}\end{eqnarray}
where $\langle 0^{\rm out}_{n}|\,0^{\rm in}_{n}\rangle$ is the probability amplitude for the
vacuum to remain a vacuum. The absolute probability for production of a single
electron-positron pair in a certain state is therefore
\begin{eqnarray}
|w^{\rm pair}_{i}|^{2}= t_{i}\,|\langle 0^{\rm out}_{i}|\,0^{\rm in}_{i}\rangle|^{2}\,,
\label{2.03}\end{eqnarray}
where $t_{i}$ is the relative probability~(\ref{1.47}) for production of a single electron-positron pair
in the state $i$ and $|\langle 0^{\rm out}_{i}|\,0^{\rm in}_{i}\rangle|^{2}$ is the probability
that no pairs are created in this state, or the vacuum-to-vacuum probability. This probability can be
determined consistently as follows.

First, we note that the absolute probability~(\ref{2.03}) must be equal to the mean number of pairs created
from the in-vacuum in a given state. Taking the expectation value of the number operator, we calculate
this explicitly
\begin{eqnarray}
{\bar n}_{i}\equiv\left\langle 0^{\rm in}_{i}\left|
\left(a^{\rm out}_{i}\right)^{\dagger}a^{\rm out}_{i}\right|0^{\rm in}_{i}\right\rangle =
\frac{1}{|R(\lambda)|^2}\left\langle 0^{\rm in}_{i}\left|b^{\rm in}_{i}\,\left(b^{\rm in}_{i}\right)^{\dagger}
\right|0^{\rm in}_{i}\right\rangle =
\frac{1}{|R(\lambda)|^2} = \frac{t_{i}}{r_{i}}\,,
\label{2.04}\end{eqnarray}
with the result as in Eqs.~(\ref{1.48}) and~(\ref{1.49}), where the transmission coefficient $t_i$
and the reflection  coefficient $r_i$ are both referred to a certain state $i$.
From Eqs.~(\ref{2.03}) and~(\ref{2.04}) it follows that
\begin{eqnarray}
|\langle 0^{\rm out}_{i}|\,0^{\rm in}_{i}\rangle|^{2} = 1/r_{i}\,.
\label{2.05}\end{eqnarray}

Second, we use the probability conservation in a given state together with the Pauli principle
for fermions to obtain directly,
\begin{eqnarray}
\left(1 + t_{i}\right)\,|\langle 0^{\rm out}_{i}|\,0^{\rm in}_{i}\rangle|^{2}
= r_{i}\,|\langle 0^{\rm out}_{i}|\,0^{\rm in}_{i}\rangle|^{2} = 1\,,
\label{2.06}\end{eqnarray}
where we apply Eq.~(\ref{1.42}) for supercritical potentials. This defines the vacuum-to-vacuum probability
$|\langle 0^{\rm out}_{i}|\,0^{\rm in}_{i}\rangle|^{2}$ as in Eq.~(\ref{2.05}).

Finally, we calculate the absolute probability amplitude for elastic scattering in a given state $n$
as follows
\begin{eqnarray}
w^{\rm scatt}_{n}&\equiv&\left\langle 0^{\rm out}_{n}\left|a^{\rm out}_{n}\left(a^{\rm in}_{n}\right)^{\dagger}\right|0^{\rm in}_{n}\right\rangle\nonumber\\
&=&-\frac{I^{\ast}(\lambda)}{R^{\ast}(\lambda)}\left\langle 0^{\rm out}_{n}\left|a^{\rm out}_{n}\left(a^{\rm out}_{n}\right)^{\dagger}\right|0^{\rm in}_{n}\right\rangle
-\frac{1}{R^{\ast}(\lambda)}\left\langle 0^{\rm out}_{n}\left|a^{\rm out}_{n}\,b^{\rm out}_{n}\right|0^{\rm in}_{n}\right\rangle\nonumber\\
&=&-\frac{I^{\ast}(\lambda)}{R^{\ast}(\lambda)}\left\langle 0^{\rm out}_{n}\left|\,
0^{\rm in}_{n}\right.\right\rangle
+\frac{1}{R^{\ast}(\lambda)}\,w^{\rm pair}_{n} =-\frac{R(\lambda)}{I(\lambda)}\left\langle 0^{\rm out}_{n}\left|\,0^{\rm in}_{n}\right.\right\rangle\,,
\label{2.07}\end{eqnarray}
where we use Eq.~(\ref{2.02}) and the current conservation~(\ref{1.39}) for supercritical potentials.
With Eq.~(\ref{1.40}), the absolute probability for elastic scattering of electron in a certain state
is therefore
\begin{eqnarray}
|w^{\rm scatt}_{i}|^{2}= r_{i}\,|\langle 0^{\rm out}_{i}|\,0^{\rm in}_{i}\rangle|^{2} =1\,,
\label{2.08}\end{eqnarray}
where the total reflection of electron by supercritical potentials is due to the Pauli principle since the initial state is already occupied.

We conclude therefore that all Eqs.~(\ref{2.05}),~(\ref{2.06}) and~(\ref{2.08}) define consistently the vacuum-to-vacuum probability for
a certain $i$-state as $1/r_{i}$, where $r_{i}$ is the reflection coefficient~(\ref{1.40}) referred to this state. For all these states,
the total vacuum-to-vacuum probability is
\begin{eqnarray}
|\langle 0^{\rm out}|0^{\rm in}\rangle|^2 = \prod_{i} (r_{i})^{-1} = \exp\left(- \sum_{i}\ln r_{i}\right) \,,
\label{2.1}\end{eqnarray}
where the products and the sums are taken over all states $i=(p,\,\sigma)$ with the quantum numbers
$p=\{\varepsilon,\,\bf{p_{\perp}}\}$ and also with the two spin projections $\sigma=1,2$.
Since $|\langle 0^{\rm out}|0^{\rm in}\rangle|^2 <1$, the vacuum is unstable producing particle-antiparticle
pairs under the influence of supercritical potentials.

The total probability for vacuum decay via pair creation reads
\begin{eqnarray}
W = 1 - \exp\left(- \sum_{i}\ln r_{i}\right)\simeq
\sum_{i}\ln r_{i} =
2\sum_{p}\ln r_{i} \,,
\label{2.2}\end{eqnarray}
where the factor $2$ is due to summation over two spin projections $\sigma =1,2$.
The remaining sum over $p=\{\varepsilon,\,\bf{p_{\perp}}\}$ can be rewritten as the integral
in a box-like volume $V_\perp T$ with $V_{\perp} = \int d^2 x_{\perp}= \int dx_1dx_2$
being the area transverse to the $z$-direction, and $T$ the total time.
This yields
\begin{eqnarray}
\sum_{p}\ln r_{i} = V_{\perp}T\,\int\frac{d^2 p_{\perp}}{(2\pi\hbar)^2}\,
\int\frac{d\,\varepsilon}{(2\pi\hbar)}\,\ln r (\varepsilon,\,p_{\perp})\,,
\label{2.3}\end{eqnarray}
where $r=r(\varepsilon,\,p_{\perp})$ with $p_{\perp}\equiv |\bf{p_{\perp}}|$
being the rotationally invariant function as it follows from Eq.~(\ref{1.40}).
This invariance reduces the integral $\int d^2 p_{\perp}$ to $\pi\int d p_{\perp}^2$.
For supercritical potentials with $\ev > m$, integrating over $\varepsilon$
and $p_{\perp}^2$ has to be done over the Klein region
\begin{eqnarray}
p_0^{(-)} = \varepsilon + v \geq \sqrt{p^{2}_\perp + m^2}\,,\quad\quad
p_0^{(+)} = \varepsilon - v \leq - \sqrt{p^{2}_\perp + m^2}\,.
\label{2.4}\end{eqnarray}

The total probability is now completely defined by Eqs.~(\ref{2.2})--(\ref{2.4}).
The vacuum decay rate (the total probability per unit cross-sectional area and unit time) 
is represented due to Nikishov~\cite{nikishov1,nikishov2} as
\begin{eqnarray}
w_{\perp} = \frac{1}{(2\pi)^2 \hbar^3}\,\int^{(\ev ^2 -m^2)}_{0}\,d p_{\perp}^2
\,\int^{\ev - \sqrt{p^{2}_\perp + m^2}}_{- \ev   + \sqrt{p^{2}_\perp + m^2}}\,d \varepsilon\,
\ln r (\varepsilon,\,p_{\perp}^2)\,,
\label{2.5}\end{eqnarray}
where the integration region in the $(p_{\perp}^2,\varepsilon)$-plane is shown in Fig.~1.
\begin{figure}[t]
\quad\quad\quad
\quad
\quad
\quad\quad\quad\quad\quad
\quad\quad
\includegraphics[width=55mm,angle=0]{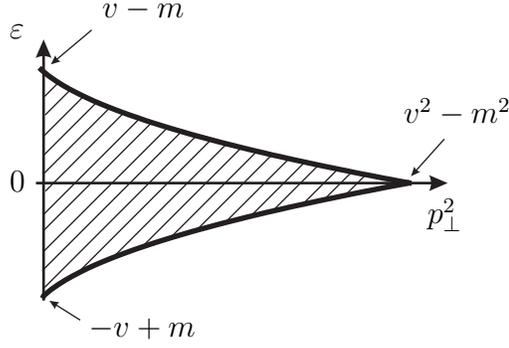}
\begin{picture}(0,10)
\put(-170,107){${\varepsilon}$}
\put(-170,49){$0$}
\put(-12,37){${p}^2_\perp$}
\put(-135,115){$v-m$}
\put(-140,-6){$-v+m$}
\put(-21,75){$v^2-m^2$}
\end{picture}
\caption{\label{Fig1} In the $(p_{\perp}^2,\varepsilon)$-plane, the integration covers
the positive region restricted by two intersecting parabolas $\varepsilon = \ev - \sqrt{p^{2}_\perp + m^2}$
and $\varepsilon = - \ev + \sqrt{p^{2}_\perp + m^2}$ with horizontal axes of symmetry
above and below the $p_{\perp}^2$-axis for $\ev > m$.}
\end{figure}
\comment{
In Eq.~(\ref{2.5}), $L\equiv V/V_{\perp}$, where $V$ is the volume occupied by field, and $V_{\perp}$
is the largest area of its cross section. For the Sauter electric field, the cross section forms
a disk of the radius $E (z)$, whose maximal value is $E_0$. So that, $V_{\perp} =\pi E_{0}^{2}$,
and the volume is
\begin{equation}
V = \pi\int_{-\infty}^\infty dz\, E^2(z) = (4\pi/3k)E^2_{0}.
\label{vol}\end{equation}
This yields
\begin{equation}
L = V/V_{\perp} = 4/3k.
\label{len}\end{equation}}
The reflection coefficient $r (\varepsilon, p_{\perp}^2)$ is defined by the symmetric under 
the interchanging $\mu\leftrightarrow\nu$ expression~(\ref{1.40}), and therefore remains invariant upon
replacing $\varepsilon\rightarrow -\varepsilon$ which interchanges merely $\mu$ and $\nu$ given by
Eqs.~(\ref{1.1b}) and~(\ref{1.12}). This allows us to obtain from (\ref{2.5}) yet another representation 
for the vacuum decay rate as follows. We interchange the order of integration
\begin{eqnarray}
w_{\perp} = \frac{1}{(2\pi)^2\hbar^3}
\left[\!\int^{0}_{-\ev + m}\!\!\!d \varepsilon\int^{(\varepsilon + \ev)^2 -m^2}_{0}\!\!\!\!
d p_{\perp}^2+\int^{\ev - m}_{0}\!\!\!d \varepsilon\int^{(\varepsilon - \ev)^2 -m^2}_{0}\!\!\!\!
d p_{\perp}^2\right]\ln r (\varepsilon, p_{\perp}^2)\,
\label{2.6}\end{eqnarray}
and replace $\varepsilon\rightarrow -\varepsilon$ in the first term of Eq.~(\ref{2.6}). Due to
the symmetry $r (-\varepsilon, p_{\perp}^2) = r (\varepsilon, p_{\perp}^2)$, the two terms in Eq.~(\ref{2.6})
are then combined into a simple integral representation
\begin{eqnarray}
w_{\perp} = \frac{1}{2\pi^2 \hbar^3}\,
\int^{\ev  -m}_{0}\,d \varepsilon\,\int^{(\varepsilon - \ev)^2 -m^2}_{0}\,d p_\perp^2 \,
\ln r (\varepsilon, p_{\perp}^2)\,,
\label{2.7}\end{eqnarray}
where the integraion region in the $(\varepsilon, p_\perp^2)$-plane is shown in Fig.~2.
\begin{figure}[b]
\quad\quad\quad\quad\quad\quad\quad\quad\quad\quad\quad
\includegraphics[width=35mm,angle=0]{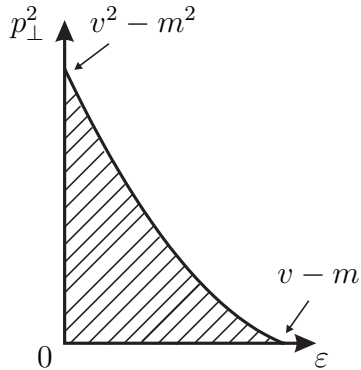}
\begin{picture}(0,12)
\put(-120,122){${p}^2_\perp$}
\put(-110,-5){$0$}
\put(-5,-5){${\varepsilon}$}
\put(-19,27){$v-m$}
\put(-90,122){$v^2-m^2$}
\end{picture}
\caption{\label{Fig2} In the $(\varepsilon,p_\perp^2)$-plane, the integration covers the region under the
left branch of the parabola \mbox{$p_\perp^2=(\varepsilon - \ev)^2 - m^2$} in the first quadrant for $\ev  >m$.}
\end{figure}
In the same way, we express also the total mean number of created electron-positron pairs
\begin{eqnarray}
{\bar N}\equiv\sum_{i}{\bar n}_{i} = 2\sum_{p}{\bar n}_{p}\,,
\label{2.10}\end{eqnarray}
where the average number of electron-positron pairs ${\bar n}_{i}$ created in a certain $i$-state
is defined by Eq.~(\ref{1.48}) independently on the spin state $\sigma =1,2$. Like the reflection
probability, it is the rotationally invariant function ${\bar n}={\bar n}(\varepsilon,\,p_{\perp})$ with the
same symmetry ${\bar n} (-\varepsilon, p_{\perp}^2) = {\bar n} (\varepsilon, p_{\perp}^2)$ under the interchanging
$\mu\leftrightarrow\nu$ in Eq.~(\ref{1.48}). The corresponding integral representation is therefore obtained
as before via replacing the sum over $p=\{\varepsilon,\,\bf{p_{\perp}}\}$ in Eqs.~(\ref{2.10}) by the integral
over the Klein region. This yields the mean number of electron-positron pairs produced by the supercritical
Sauter potential~(\ref{0.2}) from the vacuum in unit area per unit time, i.e., {\em the pair production rate},
\begin{eqnarray}
{\bar n}_{\perp} = \frac{1}{2\pi^2 \hbar^3}\,
\int^{\ev  -m}_{0}\,d \varepsilon\,\int^{(\varepsilon - \ev)^2 -m^2}_{0}\,d p_\perp^2
\,{\bar n} (\varepsilon,\,p_{\perp}^2)\,,
\label{2.11}\end{eqnarray}
with the integration region in the $(\varepsilon, p_\perp^2)$-plane shown in Fig.~2.

Eqs.~(\ref{2.7}) and~(\ref{2.11}) allow us to calculate the pair production rates
for supercritical potentials~(\ref{0.2}) with $\ev > m$. In the limit $k\rightarrow 0$ and
$v\rightarrow\infty$ (the linear potential due to constant electric field),
the obtained results can be compared with the Schwinger formula~(\ref{0.1}), and in the regime
close to this limit, with semiclassical expressions of Refs.~\cite{schub1,gies1,schub2,kim2,kim2a,kl1}.
The opposite limit $k\rightarrow\infty$ (the step potential due to delta-shaped electric field)
is however beyond the semiclassical approximation. It leads to infinite results caused by the extremely strong electric
field at the origin~\cite{grib}. In the regime close to this limit, no analytic results for production rate
were obtained so far to the best of our knowledge.

In equation~(\ref{2.7}), the reflection coefficient is given by Eq.~(\ref{1.40}),
where $\mu = \mu (\varepsilon, p_{\perp}^2)$ and $\nu = \nu (\varepsilon ,p_{\perp}^2)$
are the functions of $\varepsilon$ and $p_{\perp}^2$ defined by Eqs.~(\ref{1.12}), (\ref{1.1b})
and (\ref{1.1a}) with the constraint~(\ref{1.13}), while $\lambda = v/k\hbar >0$ is a constant.
Taking logarithms of Eq.~(\ref{1.40}) leads to the expansion
\begin{eqnarray}
\ln r  = 4\sum_{n=1}^{\infty}\,\frac{1}{n}\,\cosh(2\pi n\lambda)
\sinh(2\pi n\mu)\sinh(2\pi n\nu)\,,
\label{3.1}\end{eqnarray}
where the right hand side is found by replacing each logarithm containing the ratio
of hyperbolic functions as
$\ln(\sinh x/\sinh y) = -\sum_{n=1}^{\infty}\cosh(2nx)/n +\sum_{n=1}^{\infty}\cosh(2ny)/n$,
and combining all sums.

With Eq.~(\ref{3.1}), the vacuum decay rate~(\ref{2.7}) takes the form
\begin{eqnarray}
w_{\perp} = \frac{2}{\pi^2 \hbar^3}\sum_{n=1}^{\infty}\,\frac{1}{n}\,\cosh(2\pi n\lambda)\,J^{(n)}\,,
\label{3.2}\end{eqnarray}
where $J^{(n)}$ are the integrals
\begin{eqnarray}
J^{(n)}\! = \int^{\ev - m}_{0}\! d \varepsilon\,I^{(n)}(\varepsilon)\equiv
\int^{\ev - m}_{0}\! d \varepsilon \int^{(\varepsilon - \ev)^2 -m^2}_{0}\!d p_\perp^2\,
\sinh 2\pi n\mu (\varepsilon, p_{\perp}^2)\,\sinh 2\pi n\nu (\varepsilon ,p_{\perp}^2)\,,
\label{3.3}\end{eqnarray}
with the integration region shown in Fig.~2.

The $p_\perp^2$-integration in Eq.~(\ref{3.3}) can be done exactly~\cite{cherv}.
The result is expressed in terms of the total energy $\varepsilon$ with the help of the functions
\begin{eqnarray}
\theta_{\pm}(\varepsilon)\equiv\mu (\varepsilon, 0)\pm \nu (\varepsilon, 0)\,,\quad
\theta_{+}(\varepsilon)\,\theta_{-}(\varepsilon) = (\ev \varepsilon)/(\hbar k)^2\,,
\label{3.4}\end{eqnarray}
where $\mu (\varepsilon, 0)$ and $\nu (\varepsilon, 0)$ are two positive square roots
\begin{eqnarray}
\mu (\varepsilon,0) = \sqrt{(\varepsilon + \ev)^2 - m^2}/2\hbar k\,,\quad
\nu (\varepsilon, 0) = \sqrt{(\varepsilon - \ev)^2 - m^2}/2\hbar k\,,
\label{3.5}\end{eqnarray}
as in Eqs.~(\ref{1.12}), (\ref{1.1b}) and (\ref{1.1a}) but now with $p_\perp^2 = 0$.
Using these functions, we obtain
\begin{eqnarray}
I^{(n)}(\varepsilon) = F^{(n)} (\theta_{+}(\varepsilon)) - F^{(n)} (\theta_{-}(\varepsilon))\,,
\label{3.6}\end{eqnarray}
where
\begin{eqnarray}
F^{(n)} (\theta)&=&\left(\frac{\hbar k}{2\pi n}\right)^{2}F^{(n)}_{1} (\theta)
-\frac{(\ev\varepsilon)^2}{2}\left(\frac{2\pi n}{\hbar k}\right)^{2}F^{(n)}_{2} (\theta)\,,
\label{3.7}\end{eqnarray}
with
\begin{eqnarray}
F^{(n)}_{1} (\theta)&\equiv& {2\pi n}\,\theta\,\sinh\left({2\pi n}\,\theta\right)
- \cosh\left({2\pi n}\,\theta\right)\,,\label{3.8}\\
F^{(n)}_{2} (\theta)&\equiv& {\rm Chi}\left({2\pi n}\,\theta\right)
-\frac{2\pi n\theta\,\sinh\left({2\pi n}\,\theta\right) + \cosh\left({2\pi n}\,\theta\right)
}{\left({2\pi n}\,\theta\right)^2}\,,
\label{3.9}\end{eqnarray}
where ${\rm Chi}\left({2\pi n}\,\theta\right)$ is the hyperbolic cosine integral and the next two terms
represent the leading terms of its asymptotic expansion for large arguments ${2\pi n}\,\theta$, $n=1,2,\dots\,$.

Now the remaining $\varepsilon$-integral in Eq.~(\ref{3.3}) reads
\begin{eqnarray}
J^{(n)} = \int^{\ev - m}_{0} d \varepsilon\,I^{(n)}(\varepsilon) = J^{(n)}_{+} - J^{(n)}_{-}\,,
\label{3.10}\end{eqnarray}
where
\begin{eqnarray}
J^{(n)}_{\pm} = \int^{\ev - m}_{0} d \varepsilon\,F^{(n)} (\theta_{\pm}(\varepsilon))\,.
\label{3.11}\end{eqnarray}
The two terms in Eq.~(\ref{3.10}) can be combined into a single integral by subjecting $J^{(n)}_{\pm}$ in Eq.~(\ref{3.11})
to the following change of variable
\begin{eqnarray}
\varepsilon (\theta) = \hbar k\theta\left\{1 - \frac{m^2}{\left[\ev ^2 - \ktheta^2\right]}\right\}^{1/2}\,,
\label{3.12}\end{eqnarray}
with $0\leq \theta\leq \sqrt{\ev ^2 - m^2}/\hbar k$, where the endpoints are the zeros of the function
$\varepsilon (\theta)$, whereas its maximum $(\ev-m)$ lies at $\sqrt{\ev (\ev -m)}/\hbar k$. Note that
\mbox{$m^2\leq\left[\ev ^2 - \ktheta^2\right]$} within these limits.

For $0\leq \theta\leq\sqrt{\ev(\ev -m)}/\hbar k$, where the function~(\ref{3.12}) increases from $0$
to $(\ev-m)$, the positive square roots~(\ref{3.5}) are expressed in terms of $\theta$ as follows
\begin{eqnarray}
\sqrt{(\varepsilon\pm\ev)^2 - m^2} = \pm \hbar k\theta + v\left\{1 - \frac{m^2}{\left[\ev ^2 - \ktheta ^2\right]}\right\}^{1/2}\geq 0\,
\label{3.13}\end{eqnarray}
and therefore,
\begin{eqnarray}
\theta_{-}(\varepsilon)= \theta\,,\quad \theta_{+}(\varepsilon) =
\ev\varepsilon (\theta)/(\hbar k)^2\,\theta\,.
\label{3.14}\end{eqnarray}
Within this range, we substitute $\theta$ instead of $\varepsilon$ into the second integral
$J^{(n)}_{-}$ in Eq.~(\ref{3.11}). This yields
\begin{eqnarray}
J^{(n)}_{-} = \int^{\ev - m}_{0} d \varepsilon\,F^{(n)} (\theta_{-}(\varepsilon))
= \int^{\sqrt{\ev(\ev -m)}/\hbar k}_{0} d\theta\,\frac{d \varepsilon (\theta)}{d \theta}\,F^{(n)} (\theta)\,.
\label{3.15}\end{eqnarray}

In turn, for $\sqrt{\ev (\ev -m)}/\hbar k\leq \theta\leq \sqrt{\ev ^2 - m^2}/\hbar k$,
where the function~(\ref{3.12}) decreases from $(\ev-m)$ to $0$,
the two manifestly positive square roots~(\ref{3.5}) are
\begin{eqnarray}
\sqrt{(\varepsilon\pm\ev)^2 - m^2} = \hbar  k\theta\pm v  \left\{1 - \frac{m^2}{\left[\ev ^2 - \ktheta^2\right]}\right\}^{1/2}\geq 0\,.
\label{3.16}\end{eqnarray}
Combining these yields
\begin{eqnarray}
\theta_{+}(\varepsilon) = \theta\,,\quad \theta_{-}(\varepsilon) =
\ev\varepsilon (\theta)/(\hbar k)^2\,\theta\,.
\label{3.17}\end{eqnarray}
Within this range, $\theta$ replaces $\varepsilon$ in the first integral $J^{(n)}_{+}$ in Eq.~(\ref{3.11}).
This gives
\begin{eqnarray}
J^{(n)}_{+} = \int^{\ev - m}_{0} d \varepsilon\,F^{(n)}(\theta_{+}(\varepsilon)) =
-\int^{\sqrt{\ev^2 -m^2}/\hbar k}_{\sqrt{\ev(\ev -m)}/\hbar k} d\theta\,\frac{d \varepsilon (\theta)}
{d \theta}\,F^{(n)} (\theta)\,.
\label{3.18}\end{eqnarray}

With Eqs.~(\ref{3.18}) and~(\ref{3.15}) we obtain the integral~(\ref{3.10}) in terms of dimensionless
variable $\theta$ as follows
\begin{eqnarray}
\!\!J^{(n)}\!=\!-\!\int^{{\bar\theta}}_{0}\!d\theta\frac{d \varepsilon (\theta)}{d \theta}\,F^{(n)} (\theta)
= \int^{{\bar\theta}}_{0}\!d\theta\!\left[\!\left(\frac{\hbar k}{2\pi n}\right)^{2}\!\frac{F^{(n)}_{1} (\theta)}
{d \theta}\,\varepsilon (\theta)
-\frac{\ev^2}{6}\left(\frac{2\pi n}{\hbar k}\right)^{2}\!\frac{F^{(n)}_{2} (\theta)}
{d \theta}\,\varepsilon^{3} (\theta)\right]\!,
\label{3.19}\end{eqnarray}
where
\begin{eqnarray}
{\bar\theta}\equiv\sqrt{\ev^2 -m^2}/\hbar k\,,
\label{3.20}\end{eqnarray}
and the last term is found via partial integration thanks to the vanishing of the function~(\ref{3.12})
at the endpoints.
Substituting~(\ref{3.8}) and~(\ref{3.9}) into Eq.~(\ref{3.19}) yields
\begin{eqnarray}
J^{(n)} = (\hbar k)^3\int^{{\bar\theta}}_{0} d\theta\,f (\theta)\,\cosh \left(2\pi n\,\theta\right)\,.
\label{3.21}\end{eqnarray}
Here the function $f (\theta)$ has the form
\begin{eqnarray}
f (\theta)\equiv(\hbar k)^{-2}\left[(\hbar k\theta)\,\varepsilon (\theta) - (\ev^2/3)\,
\varepsilon^{3} (\theta)/(\hbar k\theta)^3\right]\,,
\label{3.22}\end{eqnarray}
and by Eq.~(\ref{3.10}) reads explicitly,
\begin{eqnarray}
f (\theta) = \theta^2\left(\frac{{\bar\theta}^2 -{\theta}^2}{{\lambda}^2 - {\theta}^2}\right)^{1/2}\!\!-
\,\,\,\frac{\lambda^2}{3}\left(\frac{{\bar\theta}^2 - {\theta}^2}{{\lambda}^2 - {\theta}^2}\right)^{3/2}\!\!
= f(-\theta)\,.
\label{3.23}\end{eqnarray}
Note that the integral over this function vanishes $\int^{\bar\theta}_{0}\!d \theta\, f (\theta)=0 $,
so that $J^{(0)}=0$. The integrals $J^{(n)}$ are all functions of $v,m$, and $k$.

With integrals~(\ref{3.21}) the vacuum decay rate~(\ref{3.2}) becomes
\begin{eqnarray}
w_{\perp} = \frac{2 k^3}{\pi^2}\,\int^{\bar\theta}_{0}\!d \theta\, f (\theta)
\,\sum_{n=1}^{\infty}\frac{1}{n}\cosh (2\pi n\lambda) \cosh (2\pi n\theta)\,,
\label{3.24}\end{eqnarray}
where we interchange the order of summation and integration.
For $0\leq\theta\leq{\bar\theta}\leq\lambda$, the sum
\begin{eqnarray}
\sum_{n=1}^{\infty}\frac{1}{n}\cosh (2\pi n\lambda) \cosh (2\pi n\theta) =
\frac{1}{2}\sum_{n=1}^{\infty}\frac{1}{n}\left(e^{2\pi n\lambda} + e^{-2\pi n\lambda}\right)
\cosh (2\pi n\theta)\,.
\label{3.25}\end{eqnarray}
can be found as follows. The second sum in Eq.~(\ref{3.25}) has the form
\begin{eqnarray}
\sum_{n=1}^{\infty}\frac{p^n}{n}\cosh(n x) =
- \frac{1}{2}\,\ln (1 - 2p \cosh{x} + p^2)\,,\,\,\,\,p^2 < 1\,,
\label{3.26a}\end{eqnarray}
with $x = 2\pi\theta$ and $p = \exp(- 2\pi\lambda)$. Explicitly, it reads
\begin{eqnarray}
\frac{1}{2}\sum_{n=1}^{\infty}\frac{1}{n}\,e^{-2\pi n\lambda}\cosh (2\pi n\theta) =
-\frac{1}{4}\ln\left[1 - e^{- 2\pi (\lambda + \theta)}\right] -\frac{1}{4}\ln\left[1 - e^{-2\pi (\lambda - \theta)}\right]\,.
\label{3.25a}\end{eqnarray}
The first sum in Eq.~(\ref{3.25}) can formally be written as
\begin{eqnarray}
\frac{1}{2}\sum_{n=1}^{\infty}\frac{1}{n}\,e^{2\pi n\lambda}\cosh (2\pi n\theta) =
\frac{1}{4}\,{\rm Li}_{1}(e^{2\pi (\lambda + \theta)}) + \frac{1}{4}\,{\rm Li}_{1}(e^{2\pi (\lambda - \theta)})\,,
\label{3.26b}\end{eqnarray}
where ${\rm Li}_{1}(z)$ is the polylogarithm function
\begin{eqnarray}
{\rm Li}_{\nu}(z) = \sum_{n=1}^{\infty}\,\frac{z^n}{n^{\nu}}\,,\,\,\,|z|< 1\,,
\label{3.26}\end{eqnarray}
with $\nu =1$ for which we use the analytic continuation into the region $|z|>1$,
\begin{eqnarray}
{\rm Li}_{1}\left(z\right) = {\rm Li}_{1}\left(\frac{1}{z}\right) - \ln\left(-z\right)\,.
\label{3.26c}\end{eqnarray}
Applying this to Eq.~(\ref{3.26b}), we obtain
\begin{eqnarray}
\frac{1}{2}\sum_{n=1}^{\infty}\frac{1}{n}\,e^{2\pi n\lambda}\cosh (2\pi n\theta) =
-\pi\lambda - \frac{1}{4}\,\ln\left[1 - e^{- 2\pi (\lambda + \theta)}\right] - \frac{1}{4}\,\ln\left[1 - e^{-2\pi (\lambda - \theta)}\right]\,.
\label{3.25b}\end{eqnarray}
By Eqs.~(\ref{3.25a}) and~(\ref{3.25b}), the infinite sum~(\ref{3.25}) becomes
\begin{eqnarray}
\sum_{n=1}^{\infty}\frac{1}{n}\cosh (2\pi n\lambda) \cosh (2\pi n\theta) = - \pi\lambda
-\frac{1}{2}\,\ln\left[1 - e^{- 2\pi (\lambda + \theta)}\right]  -\frac{1}{2}\,\ln\left[1 - e^{-2\pi (\lambda - \theta)}\right]\,.
\label{3.25f}\end{eqnarray}
Finally, inserting Eq.~(\ref{3.25f}) into Eq.~(\ref{3.24}) yields the vacuum decay rate
\begin{eqnarray}
w_{\perp} = - \frac{k^3}{\pi^2}\int^{\bar\theta}_{-\bar\theta}\!d \theta f (\theta)
\ln\left[1- e^{- 2\pi (\lambda - \theta)}\right],
\label{3.27}\end{eqnarray}
where we use the symmetry of the function $f(\theta) = f(-\theta)$.

The pair production rate~(\ref{2.11}) can be obtained in the same way
for large $\lambda$, i.e., close to the constant-field limit.
In this case, the average number of produced pairs~(\ref{1.48}) becomes
\begin{equation}
{\bar n} = \frac{2\sinh2\pi\mu \,\sinh2\pi\nu}{\left[\cosh2\pi\lambda - \cosh2\pi(\mu -\nu)\right]}
\simeq 4\,e^{- 2\pi\lambda}\,\sinh2\pi\mu \,\sinh2\pi\nu\,.
\label{3.29}\end{equation}
Substituting Eq.~(\ref{3.29}) into Eq.~(\ref{2.11}) yields
\begin{eqnarray}
{\bar n}_{\perp} = \frac{2}{\pi^2 \hbar^3}\,e^{- 2\pi\lambda}\, J^{(1)}
= \frac{2k^{3}}{\pi^2}\,e^{-2\pi\lambda}\,\int^{\bar\theta}_{0}\!d \theta\, f (\theta)
\,\cosh(2\pi\theta) = \frac{k^3}{\pi^2}\int^{\bar\theta}_{-\bar\theta}\!d \theta f (\theta)
e^{2\pi (\theta - \lambda)}\,,
\label{3.28}\end{eqnarray}
where the integral $J^{(1)}$ is determined by Eqs.~(\ref{3.3}) and~(\ref{3.21}) for $n=1$.
The obtained equations~(\ref{3.27}) and~(\ref{3.28}) provide us with
simple integral representations
for probabilities~(\ref{2.2}) and~(\ref{2.10}), respectively,
where the sum over all quantum numbers $\sigma$, $p_0$ and $\bf{p_{\perp}}$
is replaced by the $\theta$-integral with the function $f (\theta)$. In particular,
the pair production rate~(\ref{3.28}) is given by the first term in the expansion of logarithm in Eq.~(\ref{3.27})
for the vacuum decay rate, i.e., in the same way as Eq.~(\ref{2.10}) is related to Eq.~(\ref{2.2}).

\section{Pair production rates and field inhomogeneity}
The obtained rates~(\ref{3.27}) and~(\ref{3.28}) include the effect of spatial variations of
the electric field~(\ref{0.3}) on pair production. To make this more explicit, we represent
the function~(\ref{3.23}) in the form
$f (\theta) = (- 1/3)\partial g (\theta)/\partial\theta$ with
\begin{eqnarray}
g (\theta) = \theta\,\frac{({\bar\theta}^2 -{\theta}^2)^{3/2}}{({\lambda}^2 -{\theta}^2)^{1/2}}
= - g (-\theta)\,,
\label{4.1}\end{eqnarray}
and perform the partial integration in Eqs.~(\ref{3.27}) and~(\ref{3.28})
by taking into account that the function~(\ref{4.1}) vanishes on both ends.
Then the vacuum decay rate~(\ref{3.27}) becomes
\begin{eqnarray}
w_{\perp} = \frac{2 k^3}{3\pi}\,
\int^{\bar\theta}_{-\bar\theta}\!d \theta\,g (\theta)\,\frac{1}{e^{2\pi (\lambda - \theta)} - 1}\,.
\label{4.2}\end{eqnarray}
In what follows, we shall use the natural units in which the height $v$ and the width $1/k$
of the potential barrier~(\ref{0.2}) are measured in units of the rest energy $mc^2$ and
the Compton wavelength $\blambda_e$ of electron, respectively
\begin{eqnarray}
v \equiv\alpha\,mc^2\,,
\quad 1/k\equiv\beta\,\blambda_e\,,
\label{4.3}\end{eqnarray}
where $\alpha$ and $\beta$ are dimensionless parameters, whose ratio is
\begin{equation}
\epsilon\equiv\alpha/\beta = E_0/E_{c}\,.
\label{4.4}\end{equation}
Note that $\alpha >1$ and therefore $\beta > 1/\epsilon$ for supercritical potentials.
By these restrictions the electric field~(\ref{0.3}), whose peak $E_0$ is of the order of presently
available field strengths $\epsilon\ll 1$ can only create the pairs near the constant limit
$\beta\gg 1$, whereas for pair production near the sharp limit $\beta \ll 1$
much stronger field strengths $\epsilon\gg 1$ are required. In the units~(\ref{4.3}), the parameters~(\ref{1.14})
and~(\ref{3.20}) entered in Eq.~(\ref{4.2}) read
\begin{eqnarray}
\lambda = \alpha\beta\,,
\quad \bar\theta = \beta\sqrt{\alpha^2 -1}\,.
\label{4.5}\end{eqnarray}
Setting these up in Eq.~(\ref{4.2}) and expressing the integral in terms of
the new dimensionless variable $\vartheta =\alpha - \theta/\beta$ yields
the rate per area
\begin{eqnarray}
w_{\perp} (\alpha,\beta) = \frac{2 c}{3\pi\blambda_e^3}
\int^{\alpha + \sqrt{\alpha^2 - 1}}_{\alpha - \sqrt{\alpha^2 - 1}}
\!d \vartheta\,(\alpha - \vartheta)\,
\frac{(2\alpha\vartheta - 1 - \vartheta^2)^{3/2}}{(2\alpha\vartheta - \vartheta^2)^{1/2}}\,
\frac{\beta}{e^{2\pi\beta\vartheta} - 1}\,.
\label{4.6}
\end{eqnarray}

We define now the vacuum decay rate per volume as $w\equiv k w_{\perp} = w_{\perp}/\beta\blambda_e$.
It can be obtained by dividing both sides of Eq.~(\ref{4.6}) by the width $1/k$
of the electric field~(\ref{0.3}). This yields explicitly,
\begin{eqnarray}
w (\alpha,\beta) = \frac{2 c}{3\pi\blambda_e^4}
\int^{\alpha + \sqrt{\alpha^2 - 1}}_{\alpha - \sqrt{\alpha^2 - 1}}
\!d \vartheta\,(\alpha - \vartheta)\,
\frac{(2\alpha\vartheta - 1 - \vartheta^2)^{3/2}}{(2\alpha\vartheta - \vartheta^2)^{1/2}}\,
\frac{1}{e^{2\pi\beta\vartheta} - 1}\,.
\label{4.7}
\end{eqnarray}
The expression~(\ref{4.7}) is a function of two independent variables. Similarly to the dimensionless field~(\ref{4.4}),
it increases with growing $\alpha$ and decreases with growing $\beta$ as shown in Fig.~3.
For $\alpha$ fixed, the rate~(\ref{4.7}) vanishes in the limit $\beta\rightarrow\infty$ and diverges in the limit
$\beta\rightarrow 0$. In the later limit, the electric field~(\ref{0.3}) takes a singular $\delta$-shaped form
at the origin. The divergence of the vacuum decay rate~(\ref{4.7}) in this limit is the property particular to
the sharp step~\cite{grib}, whose discontinuity leads to enormous electric field.
\begin{figure}[t]
\centering
\hspace*{-2.3cm}
\includegraphics[width=130mm,angle=0]{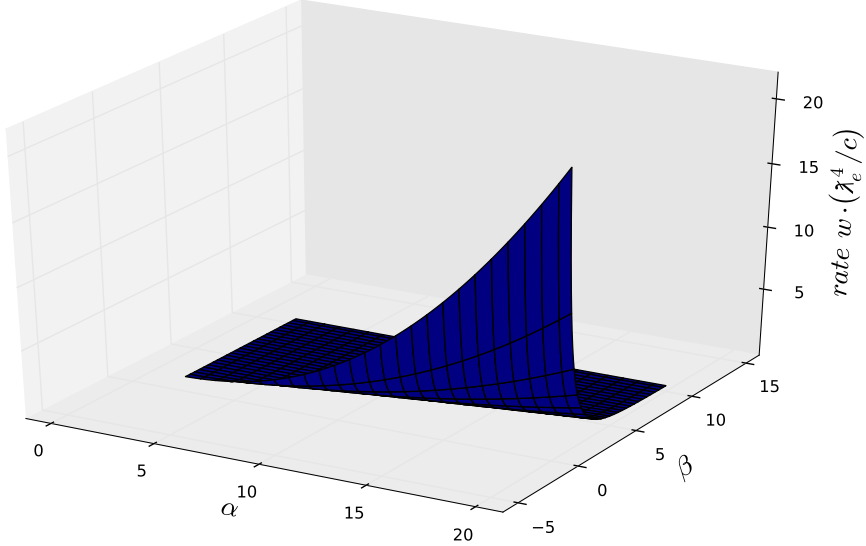}
\begin{picture}(0,10)
\end{picture}
\caption{\label{Fig3} The dimensionless rate~(\ref{4.7}) is plotted as a function of two independent
variables in natural units: the width $\beta$ and the height $\alpha$.}
\end{figure}

With increasing both arguments to infinity in such a way that their ratio $\epsilon$
remains a constant, the expression~(\ref{4.7}) becomes a homogeneous function
of this ratio. It is therefore also a constant in this limit which we denote by $w^{\rm lcf} (\epsilon)$.
The rate $w^{\rm lcf} (\epsilon)$ corresponds to a spatially uniform electric field
$\epsilon E_{c}$ obtained from the Sauter field~(\ref{0.3}) in the same limit.
For its explicit calculation, we rewrite Eq.~(\ref{4.7}) in the form
\begin{eqnarray}
&&\!\!\!\!\!\!\!\!\!\!\!\!w (\alpha,\beta)=\frac{4 c}{3\pi\blambda_e^4}\,\alpha^2\!\!
\int^{\alpha + \sqrt{\alpha^2 - 1}}_{\alpha - \sqrt{\alpha^2 - 1}}
\!d \vartheta\!\left(1-\frac{\vartheta}{\alpha}\right)\!\!
\left(\vartheta - \frac{\vartheta^2}{2\alpha}\right)^{-\frac{1}{2}}\!
\left(\vartheta - \frac{1}{2\alpha} -\frac{\vartheta^2}{2\alpha}\right)^{\frac{3}{2}}\!
\frac{1}{e^{2\pi\beta\vartheta} - 1}\nonumber\\
&&\!\!\!\!\!\!\!\!\!\!\!\!=\frac{ c}{3\pi^3 \blambda_e^4}\,\epsilon^2\!\!
\int^{2\pi\beta(\alpha + \sqrt{\alpha^2 - 1})}_{2\pi\beta(\alpha - \sqrt{\alpha^2 - 1})}
\!d \omega\!\left(1-\frac{\omega}{2\pi\alpha\beta}\right)\!\!
\left(\omega - \frac{\omega^2}{4\pi\alpha\beta}\right)^{-\frac{1}{2}}\!
\left(\omega - \frac{\pi}{\epsilon} - \frac{\omega^2}{4\pi\alpha\beta}\right)^{\frac{3}{2}}\!\!
\frac{1}{e^{\omega} - 1}\,,
\label{4.8}\end{eqnarray}
where $\omega = 2\pi\beta\vartheta$. Substituting then $\alpha = \epsilon\beta$
into Eq.~(\ref{4.8}) and taking the limit $\beta\rightarrow\infty$ with $\epsilon$ fixed, 
we obtain
\begin{eqnarray}
w^{{\rm lcf}} (\epsilon)\equiv
\left.w (\alpha\!=\!\epsilon\beta,\beta)\right|_{\beta\rightarrow\infty}
= \frac{ c}{3\pi^3 \blambda_e^4}\,\epsilon^2
\int^{\infty}_{\pi/\epsilon}\!d \omega\,\omega^{-1/2}\left(\omega - \pi/\epsilon\right)^{3/2}\,
\frac{1}{e^{\omega} - 1}\,.
\label{4.9}\end{eqnarray}
The integral~(\ref{4.9}) can be evaluated explicitly by substituting into Eq.~(\ref{4.9}) the expansion
\begin{eqnarray}
\frac{1}{e^{\omega} - 1} = \sum_{n=1}^{\infty}\,e^{- n\omega}\,
\label{4.10}\end{eqnarray}
and interchanging the order of summation and integration. In this way, we obtain
the constant-field rate per volume as
\begin{eqnarray}
w^{{\rm lcf}} (\epsilon)&=&\frac{ c}{3\pi^3 \blambda_e^4}\,\epsilon^2
\sum_{n=1}^{\infty}\int^{\infty}_{\pi/\epsilon}
\!d \omega\,\omega^{-1/2}\left(\omega - \pi/\epsilon\right)^{3/2}\,
e^{- n\omega}\nonumber\\
&=&\frac{c}{4\pi^{5/2} \blambda_e^4}\,\epsilon^2\,\sum_{n=1}^{\infty}
\frac{e^{- n\pi/\epsilon}}{n^2}\,U \left(\frac{1}{2}, -1, \frac{n\pi}{\epsilon}\right)\,,
\label{4.11}\end{eqnarray}
where $U (1/2, -1, z)$ is the Tricomi confluent hypergeometric function~\cite{abr}
with the argument $z = n\pi/\epsilon$ and the first term represents the constant-field limit of the total
mean number of created pairs~(\ref{3.28}). Thus, the constant-field limit of the vacuum decay rate~(\ref{4.7}) 
is the expression~(\ref{4.11}) rather than the Schwinger formula~(\ref{0.1}). It agrees, however, with 
formula~(\ref{0.1}), if the latter is extended to the space-dependent electric field~(\ref{0.3}) and is averaged 
over the infinite width of a spatial variation appropriated for a constant field. The expression~(\ref{4.11}) 
represents therefore the locally constant-field rate. We explain this fact and present the detailed comparison 
of the two constant-field rates in Appendix A. 
In formula~(\ref{4.11}), the ratio $\epsilon$ takes on arbitrary values despite of the conditions $\beta\gg 1$ 
and $\alpha\gg 1$. The vacuum decay rate~(\ref{4.7}) approaches the constant-field limit~(\ref{4.11}) with 
increasing $\beta$ and fixed $\epsilon$ as shown in Fig.~4.
\begin{figure}[t]
\centering
\hspace*{-2.3cm}
\includegraphics[width=210mm,angle=0]{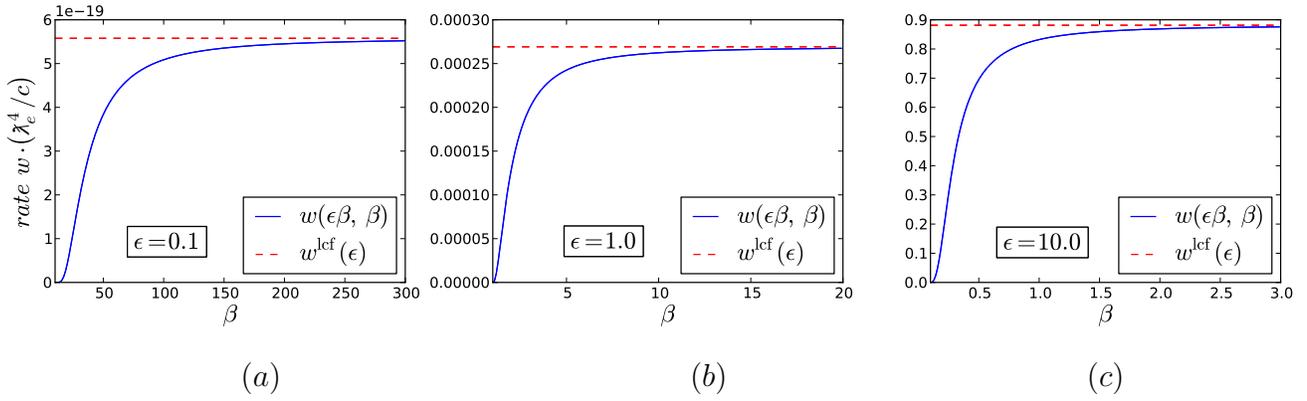}
\begin{picture}(0,10)
\put(-160,-5){$(a)$}
\put(10,-5){$(b)$}
\put(160,-5){$(c)$}
\end{picture}
\caption{\label{Fig4} The dimensionless rate~(\ref{4.7}) is plotted as a function of the field width
$\beta$ in Compton units for fixed values of the dimensionless electric field $\epsilon$ =
0.1 (a), 1.0 (b) and 10.0 (c) (solid blue lines). Dashed red lines refer to the constant-field
rate~(\ref{4.11}) for these values of the field. The variable $\beta$ starts running  from $1/\epsilon$.}
\end{figure}

Below the constant-field limit~(\ref{4.11}), the expression~(\ref{4.7}) for the vacuum decay rate per volume
interpolates analytically between the regime of sharp field $\beta \ll 1$ with $\epsilon\gg 1$ and the regime
of constant field $\beta\gg 1$ with an arbitrary $\epsilon >1/\beta$.
In order to describe such a behavior, we put the integral~(\ref{4.7})
in a more symmetric form by noting that the upper and lower limits of integration are the two zeros
of the quadratic polynomial in the numerator. By denoting these limits as
$\vartheta_{\pm}\equiv \alpha\pm\sqrt{\alpha^2 - 1}$ with $\vartheta_{+}\vartheta_{-} = 1$
and $\vartheta_{+} + \vartheta_{-} = 2\alpha$, we represent the rate~(\ref{4.7}) as follows
\begin{eqnarray}
w (\alpha,\beta) = \frac{2 c}{3\pi\blambda_e^4}
\int^{\vartheta_{+}}_{\vartheta_{-}}
\!d \vartheta\left[\left(\vartheta_{+} + \vartheta_{-}\right)/2 - \vartheta\right]
\frac{\left[\left(\vartheta_{+} - \vartheta\right)\left(\vartheta - \vartheta_{-}\right)\right]^{3/2}}
{\left[\left(\vartheta_{+} - \vartheta\right)\left(\vartheta - \vartheta_{-}\right)
+ \vartheta_{+}\vartheta_{-}\right]^{1/2}}\,\frac{1}{e^{2\pi\beta\vartheta} - 1}\,,
\label{4.18}
\end{eqnarray}
where $\vartheta_{+} > 1$ and $\vartheta_{-} < 1$. The integral~(\ref{4.18}) is dominated
by the region $\vartheta\lesssim 1/2\pi\beta$ in which the function $1/\left(e^{2\pi\beta\vartheta} - 1\right)$
differs appreciably from zero. With respect to this region, the positions of endpoints $\vartheta_{\pm}$ are
completely arbitrary. Thus locating the interval of integration as shown in Fig.~5 we can estimate the integral~(\ref{4.18}) 
analytically as follows.
\begin{figure}[t]
\centering
\hspace*{-2.2cm}
\includegraphics[width=210mm,angle=0]{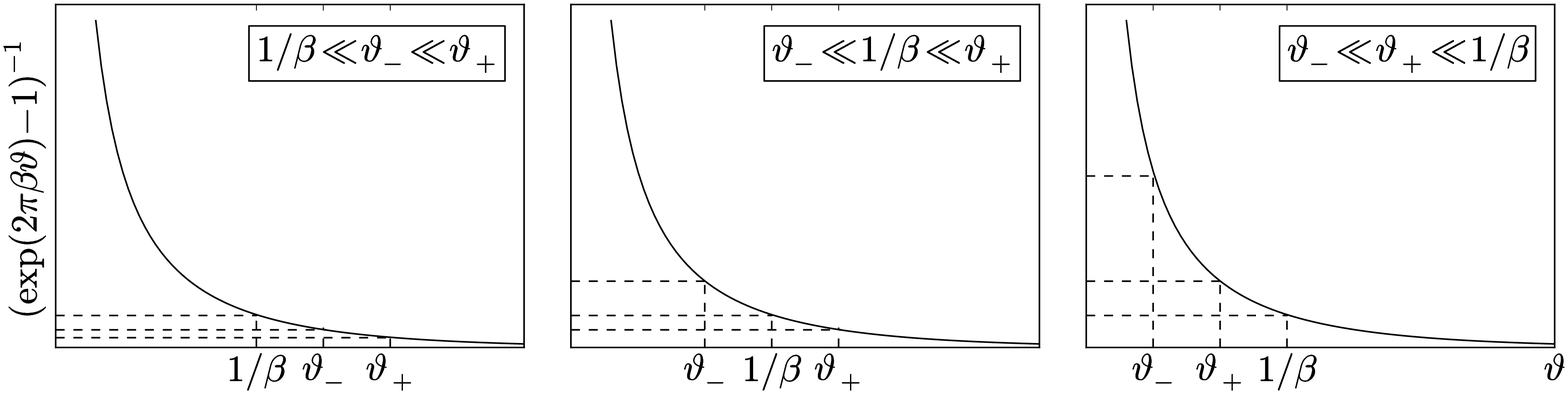}
\begin{picture}(0,10)
\put(-160,-5){$(a)$}
\put(10,-5){$(b)$}
\put(160,-5){$(c)$}
\end{picture}
\caption{\label{Fig5} The interval of integration with endpoints
$\vartheta_{\pm} = \alpha\pm \sqrt{\alpha^2 - 1}$ is shown
with respect to dominant region of the width $1/\beta$ in Eq.~(\ref{4.18}): completely outside (a),
partly outside and partly inside (b), completely inside (c). The corresponding analytic expressions
are given by Eqs.~(\ref{4.22}), (\ref{4.31}) and (\ref{4.37}), respectively.}
\end{figure}

We assume first that the position of the upper limit $\vartheta_{+}$ is far to the
right of the dominant region $\vartheta_{+}\gg 1/\beta$ as shown in Figs.~5a and~5b.
Then the region of integration is cut off by the factor $1/\left(e^{2\pi\beta\vartheta} - 1\right)$ and
is of the order $1/\beta$. In this region, $\vartheta\ll \vartheta_{+}$ and the upper limit can be
extended to the positive infinity. Thus the rate~(\ref{4.18}) becomes approximately,
\begin{eqnarray}
w (\alpha,\beta)&
\!\!\!\!\!\!\begin{array}{cc}
&\\
\simeq&\\
\beta\vartheta_{+}\gg 1
\end{array}
\!\!\!\!\!\!&
\frac{c}{3\pi\blambda_e^4}\,\vartheta_{+}^{2}\!
\int^{\infty}_{\vartheta_{-}}\!d \vartheta\,\vartheta^{-1/2}
\left(\vartheta - \vartheta_{-}\right)^{3/2}\,\frac{1}{e^{2\pi\beta\vartheta} - 1}\nonumber\\
&=&\frac{ c}{3\pi\blambda_e^4}\,\Lambda^{-2}
\int^{\infty}_{\Lambda}\!d \omega\,\omega^{-1/2}\left(\omega - \Lambda\right)^{3/2}
\!\frac{1}{e^{\omega} - 1}\equiv w (\Lambda)\,,
\label{4.19}
\end{eqnarray}
where we replace the variable $\vartheta$ by $\omega = 2\pi\beta\vartheta$ and introduce the scaling parameter
$\Lambda\equiv 2\pi\beta\vartheta_{-}$. The integral is now performed by substituting (\ref{4.10}) into (\ref{4.19})
and interchanging the order of summation and integration. Then we obtain
\begin{eqnarray}
w (\Lambda)&=&
\frac{c}{3\pi\blambda_e^4}\,\Lambda^{-2}\sum^{\infty}_{n=1}
\int^{\infty}_{\Lambda}\!d \omega\,\omega^{-1/2}
\left(\omega - \Lambda\right)^{3/2}\,e^{-n\omega}
\nonumber\\
&=&\frac{c}{4\pi^{1/2}\blambda_e^4}\,\Lambda^{-2}
\sum_{n=1}^{\infty}
\frac{e^{- n\Lambda}}{n^2}\,U \left(1/2, -1,\,n\Lambda\right)\,,
\label{4.20}
\end{eqnarray}
where all steps of calculation are similar to those in Eqs.~(\ref{4.9})-(\ref{4.11}).
This leads to the same hypergeometric function $U (1/2, -1, z)$ but now with
the argument $z = n\Lambda$. As a such, the parameter $\Lambda$ depends on $\alpha$ and $\beta$.
It can also be expressed in terms of $\alpha$ and $\epsilon$ with the help of Eq.~(\ref{4.4}).
This yields $\Lambda\equiv\left(\pi/\epsilon\right)\left(2\alpha/\vartheta_{+}\right)$ with
$\Lambda >\left(\pi/\epsilon\right)$. Thus, the rate~(\ref{4.20}) is always smaller than
its constant-filed limit~(\ref{4.11}). The latter is attained by Eq.~(\ref{4.20})
in the limit $\beta\rightarrow\infty$ with fixed $\epsilon$ via substituting $\Lambda\rightarrow\pi/\epsilon$.

In going from Eq.~(\ref{4.18}) to Eq.~(\ref{4.20}) the position of the lower limit of integration 
$\vartheta_{-}$ is still arbitrary.
For its location there are two possibilities: far to the right $\vartheta_{-}\gg 1/\beta$ (Fig.~5a)
and far to the left $\vartheta_{-}\ll 1/\beta$ (Fig.~5b) of the boundary $1/\beta$. For the first,
the interval of integration is located completely outside the dominant region, whereas for the second,
it is partly outside and mostly inside. The both are the limiting cases of Eq.~(\ref{4.20}) and
can be estimated as follows.

If the interval of integration is located completely outside the dominant region (Fig.~5a),
the rate~(\ref{4.20}) is obviously very small. The corresponding restriction $\vartheta_{-}\gg 1/\beta$ or
$\beta\vartheta_{-}\gg 1$ leads to $\beta\gg\vartheta_{+}>\alpha>1>\vartheta_{-}$. This
implies, in turn, that the initially imposed restriction $\vartheta_{+}\gg 1/\beta$ is already preserved,
and also that the electric field~(\ref{0.3}) is near its constant limit $\beta\gg 1$ with $\beta\gg\alpha$ or
$\epsilon\ll 1$. As long as $\Lambda\gg 1$, we estimate the rate~(\ref{4.20}) by making
use of the asymptotic expansion of the function $U (1/2, -1, z)$ for large arguments $z = n\Lambda$~\cite{abr}:
\begin{eqnarray}
U (1/2, -1, z)\simeq z^{-1/2}\left[1 - (5/4) z^{-1} + {\cal O}\left(z^{-2}\right)\right]\,,
\,\,\, z\rightarrow\infty\,.
\label{4.21}\end{eqnarray}
Inserting this into Eq.~(\ref{4.20}) and subjecting further to $\Lambda\gg 1$ yields
\begin{eqnarray}
w (\Lambda)&
\!\!\!\!\!\!\!\!\begin{array}{cc}
&\\
\simeq&\\
\Lambda\gg 1
\end{array}
\!\!\!\!\!\!\!\!&
\frac{c}{4\pi^{1/2}\blambda^4_{e}}
\,\Lambda^{-5/2}\left[{\rm Li}_{\frac{5}{2}}(e^{-\Lambda}) -
\frac{5}{4}\,\Lambda^{-1}\,{\rm Li}_{\frac{7}{2}}(e^{-\Lambda})+
\cdots\right]
\nonumber\\
&
\!\!\!\!\!\!\!\!\begin{array}{cc}
&\\
\simeq&\\
\Lambda\gg 1
\end{array}
\!\!\!\!\!\!\!\!&
\frac{c}{4\pi^{1/2}\blambda^4_{e}}
\,\Lambda^{-5/2}\exp\left[-\Lambda
\left(1 + \frac{5}{4}\,\Lambda^{-2} + {\cal O}\left(\Lambda^{-3}\right) \right)\right]\,.
\label{4.22}
\end{eqnarray}
The rate~(\ref{4.22}) describes the pair production in the weak-field regime $\epsilon\ll 1$ near the
constant limit $\beta\gg 1$ with $1<\alpha\ll\beta$. In terms of $\alpha$ and $\epsilon$, it can be represented  
via the substitution $\Lambda\equiv\left(\pi/\epsilon\right)\left(2\alpha/\vartheta_{+}\right)$ as
\begin{eqnarray}
w (\alpha,\epsilon)
\!\!\!\!\!\begin{array}{cc}
&\\
\simeq&\\
\epsilon\ll\alpha/\vartheta_{+}
\end{array}
\!\!\!\!\!\!
\frac{c}{4\pi^{3}\blambda^4_{e}}\,\epsilon^{5/2}
\left(\frac{\vartheta_{+}}{2\alpha}\right)^{5/2}
\exp\left\{-\frac{\pi}{\epsilon}\frac{2\alpha}{\vartheta_{+}}
\left[1 + \frac{5}{4}\left(\frac{\epsilon}{\pi}\frac{\vartheta_{+}}{2\alpha}\right)^2
- \cdots\right]\right\}\,,
\label{4.23}
\end{eqnarray}
where $\epsilon\ll\alpha/\vartheta_{+}< 1$. The leading term of the expansion~(\ref{4.23})
with slightly different pre-exponential factor including $\sqrt{\alpha^2 - 1}/\alpha$ instead of 
$\vartheta_{+}/2\alpha$ was found in Ref.~\cite{kim2} by integrating the Nikishov result and also 
in Refs.~\cite{schub1,schub2,kim2,kl1} by various semiclassical approximations.~\footnote{Note that 
our parameter $\alpha = v/mc^2 >1$ coincides with $1/\tilde{\gamma}$
of~\cite{schub1,schub2}, with $1/\epsilon$ of~\cite{kim2,kim2a} and with $\sigma$ of~\cite{kl1}.}
The corrections in powers of small $\epsilon$ were obtained in Ref.~\cite{kim2a}.

In the limit $\beta\rightarrow\infty$ with fixed $\epsilon\ll 1$ we obtain from (\ref{4.22}) 
the small-$\epsilon$ asymptotic~(\ref{a.18}) of the constant-field rate~(\ref{4.11}) by replacing 
$\Lambda\rightarrow \pi/\epsilon\gg 1$. With increasing $\alpha$ from moderate $\alpha>1$ to large 
$\alpha\gg 1$ values until $\alpha\ll\beta$, the rate~(\ref{4.22}) increases approaching the asymptotic~(\ref{a.18}) 
from below. For $\alpha^3\gg\beta$, the corrections can be obtained by expanding the scaling parameter 
$\Lambda\simeq (\pi/\epsilon)\left(1+ 1/4\alpha^2+\cdots\right)$ and then the expression~(\ref{4.22}) as follows
\begin{eqnarray}
w (\alpha,\epsilon)
\!\!\!\!\!\begin{array}{cc}
&\\
\simeq&\\
\epsilon\ll 1
\end{array}
\!\!\!\!\!\!
\frac{c}{4\pi^{3}\blambda^4_{e}}
\,\epsilon^{5/2}\exp\left\{-\frac{\pi}{\epsilon}
\left[1 + \frac{5}{4}\frac{\epsilon^2}{\pi^2} +
\frac{1}{4\alpha^{2}} + {\cal O}\left(\frac{1}{\epsilon^2\alpha^{4}}\right)\right]\right\}\,,
\quad\alpha\ll\beta\ll\alpha^3\,.
\label{4.22a}
\end{eqnarray}

If the interval of integration is located only partly outside the dominant region while occupying the most
of its part inside as shown in Fig.~5b, the rate~(\ref{4.20}) becomes very large. The corresponding 
restriction $\vartheta_{-}\ll 1/\beta$ together with the initial restriction $\vartheta_{+}\gg 1/\beta$ 
leads to the condition $\alpha\gg 1$ for which the two restrictions are reduced to $\alpha\beta\gg 1$ 
and $\beta/\alpha\ll 1$, or $\epsilon\gg 1$. As long as $\Lambda\ll 1$, estimating the rate~(\ref{4.20}) 
becomes more involved than in the previous case. Indeed, the use of the asymptotic expansion of the function 
$U (1/2, -1, z)$ for small arguments~\cite{abr}:
\begin{eqnarray} 
U (1/2, -1, z)\simeq\frac{4}{3\sqrt{\pi}}\left(1 - \frac{z}{2} + {\cal O}\left(z^{2}\right)\right)\,,
\,\,\, z\rightarrow 0\,
\label{4.24}\end{eqnarray}
with $z = n\Lambda$ leads to a slow convergent series for large $n$ in Eq.~(\ref{4.20}). 
Therefore we determine the small-$\Lambda$ behavior of the rate~(\ref{4.20})
from the integral representation~(\ref{4.19}) as follows. First of all, we rewrite this in the form
\begin{eqnarray}
w (\Lambda) =
\frac{ c}{3\pi\blambda_e^4}\,\Lambda^{-2}
\int^{\infty}_{\Lambda}\!d \omega\left(1 - \frac{\Lambda}{\omega}\right)^{3/2}
\!\frac{\omega}{e^{\omega} - 1}\,.
\label{4.25}\end{eqnarray}
In order to perform the integral~(\ref{4.25}) for $\Lambda\ll 1$, we expand
\begin{eqnarray}
\left(1 - \frac{\Lambda}{\omega}\right)^{3/2} = \sum_{n=0}^{\infty}a_{n}\Lambda^{n}\omega^{-n} =
1 - \frac{3\Lambda}{2}\omega^{-1} + \frac{3\Lambda^2}{8}\omega^{-2} + \,\cdots\,,
\label{4.26}\end{eqnarray}
with the coefficients
\begin{eqnarray}
a_n = \frac{3(2n - 1)!!}{2^{n}(2n - 1)(2n - 3)n!}\,.
\label{4.27}\end{eqnarray}
Substituting (\ref{4.26}) into (\ref{4.25}) and separating the integral into two parts yields
\begin{eqnarray}
w (\Lambda) =
\frac{ c}{3\pi\blambda_e^4}\,\Lambda^{-2}\sum_{n=0}^{\infty}a_{n}\Lambda^{n}
\left(\int^{\infty}_{0}\frac{d \omega\,\omega^{1-n}}{e^{\omega} - 1} -
\int^{\Lambda}_{0}\frac{d \omega\,\omega^{1-n}}{e^{\omega} - 1}\right)\,.
\label{4.28}\end{eqnarray}
Here the first integral is the product $\Gamma(2 - n)\zeta(2 - n)$, where $\Gamma(2 - n)$
and $\zeta(2 - n)$ are the gamma and zeta functions, respectively~\cite{abr}.
In the presence of $\Gamma(2 - n)$, each term of the infinite sum with $n\geq 2$ over this product
consists of the singular part $(-1)^{n}\zeta(2 - n)/(n - 2)! = B_{n-1}/(n - 1)!$ plus the regular part
$(-1)^{n}\psi(n - 1)\zeta(2 - n)/(n - 2)! = \psi(n - 1)B_{n-1}/(n - 1)!$, where $B_{n-1}$ are the Bernoulli
numbers and $\psi(n - 1)$ is the polygamma function~\cite{abr}. In the second integral of Eq.~(\ref{4.28})
with the small upper limit $\Lambda\ll 1$, we substitute the expansion
\begin{eqnarray}
\frac{\omega}{e^{\omega} - 1} = \sum_{k=0}^{\infty}\frac{B_k}{k!}\,\omega^{k} =
1 - \frac{\omega}{2} + \frac{\omega^2}{12} -\,\cdots\,,
\quad\omega\rightarrow 0\,,
\label{4.29}\end{eqnarray}
and extract from the double sum the logarithmically divergent term with $n = k + 1$
involving the coefficients $B_{n-1}/(n - 1)!$ after the integration. This cancels the singular part
coming from the gamma function $\Gamma(2 - n)$ to each power of small $\Lambda$.
Then, the rate~(\ref{4.28}) takes the form
\begin{eqnarray}
w (\Lambda) &=&
\frac{ c}{3\pi\blambda_e^4}\,\Lambda^{-2}\Bigg[\frac{\pi^2}{6} - \ln\Lambda
\sum_{n=1}^{\infty}\Lambda^{n}\frac{a_{n}B_{n-1}}{(n-1)!} +
\sum_{n=2}^{\infty}\Lambda^{n}\frac{a_{n}\psi(n-1)B_{n-1}}{(n-1)!}\Bigg.\nonumber\\
&-& \Bigg.
\sum_{n=0\atop n\neq k+1}^{\infty}a_{n}\sum_{k=0}^{\infty}\Lambda^{k+1}\frac{B_{k}}{(k-n+1)k!}\Bigg]\,.
\label{4.30}\end{eqnarray}
For $\Lambda\ll 1$, the expansion~(\ref{4.30}) is well approximated by the first few terms.
In doing so, we obtain finally,
\begin{eqnarray}
\!\!\!w (\Lambda)&
\!\!\!\!\!\!\!\begin{array}{cc}
&\\
\simeq&\\
\Lambda\ll 1
\end{array}
\!\!\!\!\!\!\!\!&
\frac{ c}{3\pi\blambda_e^4}\,\Lambda^{-2}\Bigg\{\frac{\pi^2}{6} + \Lambda\Bigg[\frac{3}{2}\ln\Lambda
+ S_{1}\Bigg]
+ \Lambda^{2}\Bigg[\frac{3}{16}\left(\ln\Lambda + \gamma\right)
- S_{2}\Bigg] + {\cal O}\left(\Lambda^{3}\right)\Bigg\}\,,
\label{4.31}\end{eqnarray}
where $\gamma\simeq 0.5772157\dots$ is the Euler constant~\cite{abr} and the sums $S_1$ and $S_2$ read explicitly,
\begin{eqnarray}
S_1&\equiv&\sum_{n=0\atop n\neq 1}^{\infty}\frac{a_{n}}{(n-1)} = -1 + \sum_{n=2}^{\infty}\frac{a_{n}}{(n-1)}
\simeq - 0.5794415\dots\,,\nonumber\\
S_2&\equiv&\frac{1}{2}\sum_{n=0\atop n\neq 2}^{\infty}\frac{a_{n}}{(n-2)} = -\frac{1}{4} + \frac{3}{4} +
\frac{1}{2}\sum_{n=3}^{\infty}\frac{a_{n}}{(n-2)}\simeq 0.5411802\dots\,.
\label{4.32}
\end{eqnarray}
The rate~(\ref{4.31a}) describes the pair production by extremely strong electric fields $\epsilon\gg 1$ 
within the following range of the parameters $\alpha\gg 1$  and $1/\alpha\ll\beta\ll\alpha$.
Under these conditions, the electric field~(\ref{0.3}) can either be near the sharp limit $1/\alpha\ll\beta\ll 1$, 
or near the constant limit $1\ll\beta\ll\alpha$.  Thus, the two regimes of pair production $\beta\ll 1$ and $\beta\gg 1$ 
bear a neat similarity for very large values $\alpha\gg 1$.

In the limit $\beta\rightarrow\infty$ with fixed $\epsilon\gg 1$, we obtain from (\ref{4.31}) the large-$\epsilon$
asymptotic~(\ref{a.22}) of the constant-field rate~(\ref{4.11}) via substituting $\Lambda\rightarrow \pi/\epsilon\ll 1$.
As long as $\alpha\gg 1$, the parameter $\Lambda$ in (\ref{4.31}) can be expanded as 
$\Lambda\simeq (\pi/\epsilon)\left(1+ 1/4\alpha^2+\cdots\right)$ and the expression~(\ref{4.31}) can be
then re-expanded in powers of small $1/\alpha$ as follows
\begin{eqnarray}
w (\alpha, \beta)&
\!\!\!\!\!\!\!\!\begin{array}{cc}
&\\
\simeq&\\
\alpha\gg 1
\end{array}
\!\!\!\!\!\!\!\!&
\frac{ c}{3\pi^{3}\blambda_e^4}\Bigg\{\epsilon^{2}\Bigg[\frac{\pi^2}{6}
+\frac{\pi}{\epsilon}\Bigg(\frac{3}{2}\ln\frac{\pi}{\epsilon} + S_{1}\Bigg)
+\Bigg(\frac{\pi}{\epsilon}\Bigg)^{2}\Bigg(\frac{3}{16}\ln\frac{\pi}{\epsilon} + \frac{3\gamma}{16} - S_{2}\Bigg)\Bigg]
\Bigg.\nonumber\\
&-&\Bigg.\frac{\pi^2}{12\beta^2}
-\frac{\pi}{4\alpha\beta}\Bigg(\frac{3}{2}\ln\frac{\pi}{\epsilon} - \frac{3}{2} + S_{1}\Bigg)
+ {\cal O}\left(\alpha^{-2}\right)\Bigg\}\,,\quad 1/\alpha\ll\beta\ll\alpha\,.
\label{4.31a}\end{eqnarray}
In the regime $1/\alpha\ll\beta\ll 1$, the rate~(\ref{4.31a}) is very large but is overestimated
by the large-$\epsilon$ asymptotic~(\ref{a.22}). With growing $\beta$ the expansion~(\ref{4.31a}) 
decreases approaching the asymptotic~(\ref{a.22}) in the regime $1\ll\beta\ll\alpha$. 

Let us assume now that the position of the upper limit $\vartheta_{+}$ lies far to the left
$\vartheta_{+}\ll 1/\beta$ of the boundary $1/\beta$ as shown in Fig.~5c. The restriction
$\vartheta_{+}\ll 1/\beta$ or $\beta\vartheta_{+}\ll 1$ leads to $\beta\ll\vartheta_{-}<1<\alpha<\vartheta_{+}$.
This yields, in turn, the restriction $\beta\vartheta_{-}\ll 1$ for the lower limit $\vartheta_{-}$ and
implies that the electric field~(\ref{0.3}) is near the sharp limit $\beta\ll 1$ with $\beta\ll\alpha$
or $\epsilon\gg 1$. The parameter $\alpha$ is restricted by the condition $\alpha<\vartheta_{+}\ll 1/\beta$.
The integral~(\ref{4.18}), where the interval of integration is located completely inside the dominant region 
can be estimated as follows. Within this interval, we use the representation
\begin{eqnarray}
\frac{1}{e^{2\pi\beta\vartheta} - 1}=
\sum_{k=0}^{\infty}\frac{B_k}{k!}\,\left(2\pi\beta\vartheta\right)^{k-1}
\simeq \frac{1}{2\pi\beta\vartheta}\left(1 - \frac{2\pi\beta\vartheta}{2}
+ \cdots\right)\,,
\label{4.33}\end{eqnarray}
where $B_k$ are the Bernoulli numbers~\cite{abr}. Substituting Eq.~(\ref{4.33}) into Eq.~(\ref{4.18})
and making the change of integration variable $\vartheta = \left[\left(\vartheta_{+}-\vartheta_{-}\right)\upsilon +\left(\vartheta_{+}+\vartheta_{-}\right)\right]/2$,
we obtain
\begin{eqnarray}
w (\alpha,\beta) =
\frac{c}{3\pi^2 \blambda_e^4}\,\frac{\alpha^{3}}{\beta}\sum_{k=0}^{\infty}\frac{B_k}{k!}\,
(2\pi\alpha\beta)^{k}\,I_{k}(a)\,,
\label{4.34}
\end{eqnarray}
where $a\equiv\sqrt{\alpha^2 - 1}/\alpha < 1$ and the integrals
\begin{eqnarray}
I_{k}(a) = a^{5}\!\!\int^{1}_{-1}\!d \upsilon\,\frac{\upsilon\,(1-\upsilon^2)^{3/2}}{(1 - a^{2}\upsilon^2)^{1/2}}
\,(1 - a\upsilon)^{k-1}\,
\label{4.35}
\end{eqnarray}
are certain combinations of the complete elliptic integrals ${\bf K} (a)$ and ${\bf E} (a)$
of the first and the second kind, respectively. Explicitly, these read
\begin{eqnarray}
&&I_{0}(a) = \frac{2}{3}\left[\left(8 - 7 a^2\right){\bf E}(a) - \left(8 - 11 a^2 + 3 a^4\right){\bf K}(a)\right],
\,\,I_{1}(a) = 0\,,\nonumber\\
&&I_{2}(a) = \frac{2}{15}\left[\left(8 + 5 a^2 + 3 a^4\right){\bf E}(a) -
\left(8 + 9 a^2 + 12 a^4\right){\bf K}(a)\right]\,,\dots\,.
\label{4.36}
\end{eqnarray}
For $\alpha\beta\ll 1$, retaining only the first few terms in Eq.~(\ref{4.34}) provides us with a good approximate rate.
In fact, the expression~(\ref{4.34}) represents a quickly convergent series
\begin{eqnarray}
w (\alpha, \beta)
\!\!\!\!\begin{array}{cc}
&\\
\simeq&\\
\beta\vartheta_{+}\ll 1
\end{array}
\!\!\!\!
\frac{c}{3\pi^2 \blambda_e^4}\,\frac{\alpha^{3}}{\beta}
\left[I_{0}(a) + \frac{\pi^2}{3}\,I_{2}(a)\,(\alpha\beta)^{2} + {\cal O}\left((\alpha\beta)^{3}\right)\right]
\,,\quad \alpha\beta\ll 1\,.
\label{4.37}
\end{eqnarray}
The rate~(\ref{4.37}) describes the pair production in the strong-field regime $\epsilon\gg 1$
near the sharp-field limit $\beta\ll 1$ for all values of the parameter $\alpha$ satisfying  $1<\alpha\ll 1/\beta$.
With increasing $\alpha$ from moderate $\alpha>1$ to large $\alpha\gg 1$ values until $\alpha\ll 1/\beta$,
it can be expanded further as
\begin{eqnarray}
w (\alpha,\beta)&
\!\!\!\!\!\!\!\!\begin{array}{cc}
&\\
\simeq&\\
\beta\ll 1
\end{array}
\!\!\!\!\!\!\!\!&
\frac{2 c}{9\pi^2 \blambda_e^4}\frac{\alpha^{3}}{\beta}
\left[1 + \frac{9}{2}\left(\frac{3}{2} - \log4\alpha\right)\frac{1}{\alpha^2} +
{\cal O}\left(\frac{1}{\alpha^4}\right)\right]\,,\,\, 1\ll\alpha\ll 1/\beta\,,
\label{4.37a}
\end{eqnarray}
where $\alpha^3/\beta\equiv\epsilon\alpha^2 \ll\epsilon^2$ as long as $\alpha\beta\ll 1$.
The rate~(\ref{4.37a}) is therefore smaller than the rate~(\ref{4.31a}) in the same strong-field
regime of pair production.

The pair production near the constant-field limit $\beta\gg 1$ is now described as follows.
For $\beta\gg 1>\vartheta_{-}$, the restriction $\beta\vartheta_{+}\gg 1$ is always satisfied.
For moderate values of the parameter $\alpha>1$, we obtain in addition the restriction
$\beta\vartheta_{-}\gg 1$ due to $\beta\gg 1$. This yields $\beta\gg\vartheta_{+}>\alpha>1$ implying that
the regime of pair production is just the weak-field regime $\epsilon\ll 1$ described by the rate~(\ref{4.22}).
With growing $\alpha$ from moderate $\alpha>1$ to large $\alpha\gg 1$ values until $\alpha\ll\beta$,
the first restriction $\beta\vartheta_{+}\simeq\alpha\beta\gg 1$ is satisfied automatically, whereas
the second $\beta\vartheta_{-}\simeq\beta/\alpha\gg 1$ is still preserved by the condition $1\ll\alpha\ll\beta$.
With these conditions, the weak-field regime is still maintained but is described now by the large-$\alpha$
asymptotic~(\ref{4.22a}) of the expansion~(\ref{4.22}) approaching from below the small-$\epsilon$ asymptotic~(\ref{a.18})
of the constant-field rate~(\ref{4.11}) with growing $\alpha$. Increasing $\alpha$ further as $\alpha\gg\beta\gg 1$
leads to $\beta\vartheta_{-}\simeq\beta/\alpha\ll 1$ instead of $\beta\vartheta_{-}\simeq\beta/\alpha\gg 1$.
Then the pair production undergoes the transition from weak-field $\epsilon\ll 1$ to strong-field $\epsilon\gg 1$
regime, where it described by the large-$\alpha$ expansion~(\ref{4.31a}) approaching the large-$\epsilon$
asymptotic~(\ref{a.22}) of the constant-field rate~(\ref{4.11}). The increase of pair production rate due to 
transition is of the order of two magnitudes. Near $\epsilon = 1$, the rate of pair production is interpolated 
analytically by the expression~(\ref{4.18}).

Consider now the pair production near the sharp-field limit $\beta\ll 1$. For $\beta\ll 1<\vartheta_{+}$,
the restriction $\beta\vartheta_{-}\ll 1$ is always satisfied. For moderate values of the parameter $\alpha>1$,
there is in addition the restriction $\beta\vartheta_{+}\ll 1$ due to $\beta\ll 1$. This yields $\beta\ll\vartheta_{-}< 1<\alpha$.
The regime of pair production is therefore the strong-field regime $\epsilon\gg 1$ described by the rate~(\ref{4.37}).
With increasing $\alpha$ from moderate $\alpha>1$ to large $\alpha\gg 1$ values until $\alpha\ll 1/\beta$,
the first restriction $\beta\vartheta_{-}\simeq\beta/\alpha\ll 1$ is satisfied automatically, whereas the second
$\beta\vartheta_{+}\simeq\alpha\beta\ll 1$ is still preserved by the condition $1\ll\alpha\ll 1/\beta$.
The strong-field regime $\epsilon\gg 1$ of pair production in now described by the large-$\alpha$ expansion~(\ref{4.37a}).
Increasing $\alpha$ further as $\alpha\gg 1/\beta\gg 1$ yields $\beta\vartheta_{+}\simeq\alpha\beta\gg 1$ instead of $\beta\vartheta_{+}\simeq\alpha\beta\ll 1$. Then the strong-field regime of pair production undergoes the smooth 
transition from $\epsilon\ll 1/\beta^2$ to $\epsilon\gg 1/\beta^2$, where it described by the large-$\alpha$ 
expansion~(\ref{4.31a}). 

For very large values $\alpha\gg 1$, the pair production is thus described by the expression~(\ref{4.31a}) 
whatever $\beta\gg 1$ or $\beta\ll 1$. In the strong-field regime $\epsilon\gg 1$, this expression becomes very large but
does not exceed the large-$\epsilon$ asymptotic~(\ref{a.22}) of the locally constant-field rate~(\ref{4.11}). 
The latter is therefore the upper limit for the vacuum decay rate~(\ref{4.18}). Thus, in agreement with 
Refs.~\cite{dunne4,schub1,schub2,kl1,gies2}, we come to the conclusion that the spatial variations of the Sauter 
field~(\ref{0.3}) cannot increase the pair production over the locally constant-field rate~(\ref{4.11}) even 
for extremely strong fields with $\epsilon\gg 1$. The later rate overestimates however the Schwinger formula~(\ref{a.6}) 
as long as $\epsilon\gg 1$.

\section{Conclusion}
We have calculated analytic expressions for the production rate of electron-positron pairs 
from the vacuum by the Sauter potential. For an arbitrary potential barrier, the rate was related 
to the scattering amplitude on the barrier, and expressed as an energy-momentum integral over the 
logarithm of the reflection coefficient. For the Sauter potential, we have evaluated the rotationally 
invariant integral over transverse momenta in three dimensions exactly. The remaining integral over the 
energy gave us the simple spectral formulas for the vacuum decay and pair production rate.
The analytic expressions for these rates were derived for the entire range of the parameters
$v$ and $k$ of the supercritical potential~(\ref{0.2}). This allowed us to access different 
physical regimes from weak to strong fields.

~\\
{Acknowledgement}:\\[2mm]
A.C. is grateful A.~Melnikov for many useful discussions.

\appendix
\section{Appendix: Constant-field limit}
In previous studies of pair creation by inhomogeneous fields, the transition to the constant
field has often been made for the local probabilities before the energy-momentum
integration. In order to outline this procedure, we consider
the densities for reflection and transmission probabilities, and also for the average number of pairs
created in each mode given by Eqs.~(\ref{1.40}), ~(\ref{1.47}) and~(\ref{1.48}), respectively.

The reflection and transmission coefficients in Eqs.~(\ref{1.40}) and~(\ref{1.47}) can be written as
\begin{eqnarray}
r = \frac{1}{1 - {\bar n}}\,,\quad\quad\quad t = \frac{{\bar n}}{1 - {\bar n}}\,,
\label{a.1}\end{eqnarray}
with $r - t = 1$ for supercritical potentials, where ${\bar n}$ is the density number
of produced pairs~(\ref{1.48}).  In order to find this quantity in the constant-field limit
$k\rightarrow 0$ with fixed $E_{0}$, we first substitute $v = (|e|E_{0})/k$. Then evaluating
the limit yields
\begin{eqnarray}
{\bar n} = \exp\left[-\pi\left(p^2_{\perp} + m^2\right)/(\hbar|e|E_0)\right]\,.
\label{a.2}\end{eqnarray}
By Eq.~(\ref{a.2}), the reflection coefficient in Eq.~(\ref{a.1}) becomes independent
of the energy $\varepsilon$. Its logarithm is
\begin{eqnarray}
\ln r (p_{\perp}^2) = - \ln\left(1 - {\bar n}\right) = \sum_{n=1}^{\infty}\frac{1}{n}
\exp\left[- n\pi\left(p^2_{\perp} + m^2\right)/(\hbar |e|E_0)\right]\,.
\label{a.3}\end{eqnarray}
With Eq.~(\ref{a.3}) the vacuum decay rate per area~(\ref{2.7}) in the constant-field limit
becomes
\begin{eqnarray}
w^{\rm cf}_{\perp} (E_0)&=& \frac{1}{(2\pi)^2 \hbar^3}\,\int^{+\infty}_{-\infty}\,d \varepsilon\,
\int^{\infty}_{0}\,d p_{\perp}^2\,\ln r (p_{\perp}^2)\nonumber\\
&=& \frac{1}{(2\pi)^2 \hbar^3}\,\sum_{n=1}^{\infty}\frac{1}{n}
\,\int^{+\infty}_{-\infty}\,d \varepsilon\,
\int^{\infty}_{0}\,d p_{\perp}^2\,e^{- n\pi\left(p^2_{\perp} + m^2\right)/(\hbar |e|E_0)}\,.
\label{a.4}\end{eqnarray}
Here the $p_{\perp}^2$-integral results in  a factor of $\hbar |e|E_{0}/n\pi$
and the $\varepsilon$-integral, via the substitution $d\varepsilon = |e|E_{0}\,d z$, in
a factor of $|e|E_{0}L$, where $L$ is the infinite length. This yields the constant-field rate per area
\begin{eqnarray}
w^{\rm cf}_{\perp} (E_0) = L\,w^{\rm cf} (E_0)\,,
\label{a.5}\end{eqnarray}
where $w^{\rm cf} (E_0)$ is the vacuum decay rate per volume given by the Schwinger formula~(\ref{0.1}).
For the constant electric field $E_0$, it reads
\begin{eqnarray}
w^{\rm cf} (E_0)=\frac{(e E_0)^2}{4\pi^3 \hbar^2 c}\,\sum_{n=1}^{\infty}\frac{1}{n^2}\,e^{-n\pi(E_{c}/E_0)}
=\frac{c}{4\pi^3 \blambda_{e}^4}\,\epsilon^2\,\sum_{n=1}^{\infty}\frac{1}{n^2}\,e^{-n\pi/\epsilon}
= w^{\rm cf} (\epsilon)\,,
\label{a.6}\end{eqnarray}
with $\epsilon\equiv E_{0}/E_{c}$. Thus, the finite rate per volume~(\ref{a.6}) is obtained
from the infinite rate per area~(\ref{a.5}) via dividing by the infinite factor of a length.

In order to compute the limit $k\rightarrow 0$ after the energy-momentum integration,
we modify the above derivation by noting that $E (z)\rightarrow E_0$ for $k\rightarrow 0$,
where $E (z)$ is the electric field~(\ref{0.3}).  Then Eq.~(\ref{a.3}) can be rewritten as
\begin{eqnarray}
\ln r (p_{\perp}^2) = \lim\limits_{k\to 0}\sum_{n=1}^{\infty}\frac{1}{n}
\exp\left[- n\pi\left(p^2_{\perp} + m^2\right)/(\hbar |e|E (z))\right]\,.
\label{a.7}\end{eqnarray}
The constant-field rate per area~(\ref{a.4}) becomes
\begin{eqnarray}
w^{\rm cf}_{\perp} (E_0)&=&\frac{1}{(2\pi)^2 \hbar^3}\,\lim\limits_{k\to 0}\sum_{n=1}^{\infty}\frac{1}{n}
\,\int^{+\infty}_{-\infty}\,d \varepsilon\,
\int^{\infty}_{0}\,d p_{\perp}^2\,e^{- n\pi\left(p^2_{\perp} + m^2\right)/(\hbar |e|E (z))}\nonumber\\
&=&\frac{1}{4\pi^3 \hbar^2}\,\lim\limits_{k\to 0}\sum_{n=1}^{\infty}\frac{1}{n^2}
\,\int^{+\infty}_{-\infty}\,d \varepsilon\,(|e|E (z))
e^{- n\pi(E_{c}/E (z))}\,,
\label{a.8}\end{eqnarray}
where we interchange the limit and the energy-momentum integral and perform
the integration over $p_{\perp}^2$. In the remaining integral, the energy $\varepsilon$
is related to the position $z$ of the electric field $E (z)$ as $d\varepsilon = |e|E (z)\,d z$.
For the Sauter electric field~(\ref{0.3}), this substitution yields
\begin{eqnarray}
w^{\rm cf}_{\perp} (E_0) &=& \frac{1}{4\pi^3 \hbar^2}\,\lim\limits_{k\to 0}\sum_{n=1}^{\infty}\frac{1}{n^2}
\,\int^{+\infty}_{-\infty}\,d z\,(|e|E (z))^{2}\,e^{- n\pi(E_{c}/E (z))}\nonumber\\
&=& \frac{c}{4\pi^{3}\blambda^4_{e}}\,\epsilon^{2}
\lim\limits_{k\to 0}\sum_{n=1}^{\infty}\frac{1}{n^2}\,
\int^{+\infty}_{-\infty}\,\frac{d z}{\cosh^{4}k z}\,e^{- (n\pi/\epsilon)\cosh^{2} k z}\,.
\label{a.9}\end{eqnarray}
By changing the variable $y = kz$ the constant-field rate per area~(\ref{a.9}) takes the form
\begin{eqnarray}
w^{\rm cf}_{\perp} (E_0) = \left(\lim\limits_{k\to 0}\frac{1}{k}\right)w^{\rm lcf} (E_0)\,,
\label{a.10}\end{eqnarray}
where $\lim\limits_{k\to 0}(1/k)$ plays now the role of the infinite length.
The constant-field rate per volume reads
\begin{eqnarray}
w^{\rm lcf} (E_0) &=& \frac{c}{4\pi^{3}\blambda^4_{e}}\,\epsilon^{2}
\sum_{n=1}^{\infty}\frac{1}{n^2}\,
\int^{+\infty}_{-\infty}\,\frac{d y}{\cosh^{4} y}\,e^{- (n\pi/\epsilon)\cosh^{2} y}\nonumber\\
&=&\frac{c}{4\pi^{3}\blambda^4_{e}}\,\epsilon^{2}
\sum_{n=1}^{\infty}\frac{e^{- n\pi/\epsilon}}{n^2}\,
\int^{+\infty}_{-\infty}\,\frac{d x}{(1 + x^2)^{5/2}}\,e^{-(n\pi/\epsilon)x^{2}}
= w^{\rm lcf} (\epsilon)\,,
\label{a.11}\end{eqnarray}
where $x = \sinh y$. The last integral in Eq.~(\ref{a.11}) is a certain combination of the modified
Bessel functions $K_{\nu} (z)$ with $\nu = 0,1$ and $z = n\pi/\epsilon$ which can be expressed
in terms of the function $U (1/2, -1, z)$~\cite{abr}. This yields, finally, the constant-field rate per
volume as
\begin{eqnarray}
w^{\rm lcf} (\epsilon) &=& \frac{c}{4\pi^{3}\blambda^4_{e}}\,\frac{2}{3}\,\epsilon^{2}\,
\sum_{n=1}^{\infty}\frac{e^{- n\pi/\epsilon}}{n^2}\,
\left(\frac{n\pi}{\epsilon}\right)\,e^{n\pi/2\epsilon}\left[\frac{n\pi}{\epsilon}
\,K_{0}\left(\frac{n\pi}{2\epsilon}\right) + \left(1 - \frac{n\pi}{\epsilon}\right)
K_{1}\left(\frac{n\pi}{2\epsilon}\right)\right]\nonumber\\
&=&\frac{c}{4\pi^{5/2}\blambda^4_{e}}\,\epsilon^{2}\,\sum_{n=1}^{\infty}
\frac{e^{- n\pi/\epsilon}}{n^2}\,U \left(\frac{1}{2}, -1, \frac{n\pi}{\epsilon}\right)\,.
\label{a.12}\end{eqnarray}
This result was already found in Eq.~(\ref{4.11}).
The rate~(\ref{a.12}) as well as the rate~(\ref{a.6}) corresponds to the constant electric
field $E_0$ but does not coincide with the latter. Moreover, the above derivation shows
that the Schwinger expression~(\ref{a.6}) cannot be recovered as a constant-field limit
of the non-constant field rate in which the energy-momentum integral was already performed.
On the other hand, both rates per volume are found from the same (infinite) rate per
area by the only different factorization of an infinite length and can therefore be
related to each other. Comparing (\ref{a.5}) with (\ref{a.10}), we obtain
\begin{eqnarray}
w^{\rm cf}_{\perp} (E_0) &=& \left(\lim\limits_{k\to 0}\frac{1}{k}\right)w^{\rm lcf} (E_0) =
L w^{\rm cf} (E_0) = L\lim\limits_{k\to 0} w^{\rm cf} (E (z))
\nonumber\\
&=& \lim\limits_{k\to 0}\int^{+\infty}_{-\infty}\!\! d z\,w^{\rm cf} (E (z))
= \left(\lim\limits_{k\to 0}\frac{1}{k}\right)\int^{+\infty}_{-\infty}\!\! d y\,w^{\rm cf} (E (y/k))\,.
\label{a.13}\end{eqnarray}
This expresses the rate~(\ref{a.12}) in terms of the rate~(\ref{a.6}) extended to the space-dependent
electric field~(\ref{0.3}) and averaged over the infinite width of a spatial variation
\begin{eqnarray}
w^{\rm lcf} (E_0) = \int^{+\infty}_{-\infty}\!\! d y\,w^{\rm cf} (E (y/k))
= \lim\limits_{k\to 0}k\int^{+\infty}_{-\infty}\!\! d z\,w^{\rm cf} (E (z))\,.
\label{a.14}\end{eqnarray}
The rate~(\ref{a.12}) is therefore the locally constant-field rate. The locally constant field approximation 
was introduced in Ref.~\cite{schub2}. In section~4, we have shown that the expression~(\ref{a.12}) is a 
constant-field limit of the vacuum decay rate~(\ref{4.7}).

The rate~(\ref{a.12}) possesses the integral representation~(\ref{4.9}).
Similar integral representation can also be found for
the rate~(\ref{a.6}) provided that we express the sum in Eq.~(\ref{a.6}) as
\begin{eqnarray}
{\rm Li}_{2}(e^{- \pi/\epsilon})\equiv\sum_{n=1}^{\infty}\,\frac{e^{- n\pi/\epsilon}}{n^{2}}
= \sum_{n=1}^{\infty}\int^{\infty}_{\pi/\epsilon}
\!d \theta\left(\theta - \frac{\pi}{\epsilon}\right)\,e^{- n\theta}
= \int^{\infty}_{\pi/\epsilon}
\!d \theta\left(\theta - \frac{\pi}{\epsilon}\right)\frac{1}{e^{\theta} - 1}\,.
\label{a.15}\end{eqnarray}
Inserting this into Eq.~(\ref{a.6}) yields
\begin{eqnarray}
w^{\rm cf} (\epsilon) = \frac{c}{4\pi^{3}\blambda^4_{e}}\,\epsilon^{2}\int^{\infty}_{\pi/\epsilon}
\!d \theta\left(\theta - \frac{\pi}{\epsilon}\right)\frac{1}{e^{\theta} - 1} \,.
\label{a.16}\end{eqnarray}

For better comparison of the two constant-field rates~(\ref{a.6}) and~(\ref{a.12}), let us estimate
their asymptotic behavior for small and large $\epsilon$, respectively. For weak electric fields
$\epsilon\ll 1$, a good approximation of the rate~(\ref{a.6}) is the Heisenberg-Euler expression
given by the first term of the Schwinger formula
\begin{eqnarray}
w^{\rm cf} (\epsilon) = \frac{c}{4\pi^{3}\blambda^4_{e}}\,{\epsilon}^{2}\,{\rm Li}_{2}(e^{- \pi/\epsilon})
\simeq\frac{c}{4\pi^{3}\,\blambda_{e}^{4}}\,\epsilon^2
\exp\left(-\pi/\epsilon\right)\,,\,\,\, \epsilon\ll 1\,.
\label{a.17}\end{eqnarray}
The corresponding behavior of the rate~(\ref{a.12}) can be found directly from the expansion~(\ref{4.22})
in the limit $\beta\rightarrow\infty$ with fixed $\epsilon\ll 1$. Since $\Lambda\rightarrow \pi/\epsilon\gg 1$
in this limit,  Eq.~(\ref{4.22}) takes the form
\begin{eqnarray}
w^{\rm lcf} (\epsilon)&\simeq&\frac{c}{4\pi^{3}\blambda^4_{e}}\,
\epsilon^{5/2}\left[{\rm Li}_{\frac{5}{2}}(e^{- \pi/\epsilon}) -
\left(5\epsilon/4\pi\right){\rm Li}_{\frac{7}{2}}(e^{- \pi/\epsilon})+
\cdots\right]
\nonumber\\
&\simeq&\frac{c}{4\pi^{3}\blambda^4_{e}}\,
\epsilon^{5/2}\exp\left\{\left(-\pi/\epsilon\right)\left[1 + \left(5/4\pi^2\right)\epsilon^2 +
{\cal O}\left(\epsilon^{3}\right)\right]\right\}\,,\,\,\, \epsilon\ll 1\,.
\label{a.18}\end{eqnarray}
This differs from the result~(\ref{a.17}) by the small pre-exponential factor $\sqrt{\epsilon}$
and also by small subleading terms. The Schwinger formula~(\ref{a.6}) is therefore underestimated
by the locally constant-field expression~(\ref{a.12}) for small $\epsilon$ as it shown in Fig.~6a.
Otherwise, for $n\epsilon\ll 1$, the expression under the sum sign in the first line of Eq.~(\ref{a.18})
can be represented as
\begin{eqnarray}
\frac{\epsilon^{5/2}}{n^{5/2}}\left[1 -
\frac{5}{4\pi n}\,\epsilon + \frac{105}{32\pi^2 n^2}\,\epsilon^2
\cdots\right]\exp\left(- n\pi/\epsilon\right)
\simeq\frac{\epsilon^{5/2}}{n^{5/2}}\exp\left[- n\pi/\epsilon_{*} (n)\right]
\,,\,\,\, n\epsilon\ll 1\,,
\label{a.18a}\end{eqnarray}
where
\begin{eqnarray}
\frac{1}{\epsilon_{*} (n)}\equiv\frac{1}{\epsilon} +
\frac{5}{4\pi^2 n^2}\,\epsilon - \frac{5}{2\pi^3 n^3}\,\epsilon^2\,,\,\,\, n\epsilon\ll 1\,,
\label{a.18aa}\end{eqnarray}
can be interpreted as a mass shift via the coherent production of $n$ pairs of particles~\cite{ritus2}.

For strong electric fields $\epsilon\gg 1$, the asymptotic behavior of the rate~(\ref{a.6})
can be found most easily from the integral representation~(\ref{a.15}). To this end, we separate
the last integral in Eq.~(\ref{a.15}) further into three parts
\begin{eqnarray}
{\rm Li}_{2}(e^{- \pi/\epsilon}) = \int^{\infty}_{0}\!d \theta\frac{\theta}{e^{\theta} - 1}
- \int^{\pi/\epsilon}_{0}\!d \theta\frac{\theta}{e^{\theta} - 1}
-\frac{\pi}{\epsilon}\int^{\infty}_{\pi/\epsilon}\!d \theta\frac{1}{e^{\theta} - 1}\,.
\label{a.19}\end{eqnarray}
Here the first integral is the number $\Gamma(2)\,\zeta(2)=\pi^2/6$,
the second is represented for $\theta\leq \pi/\epsilon\ll 1$ by the expansion
$\pi/\epsilon - \pi^2/4\epsilon^2 + \cdots \simeq \pi/\epsilon$, the third is equal to
$-\ln(1 - e^{-\pi/\epsilon})$ and can be expanded further as
$-\ln\pi/\epsilon + \pi/2\epsilon + \cdots\,$.
Collecting all terms together yields the expansion
\begin{eqnarray}
{\rm Li}_{2}(e^{- \pi/\epsilon}) \simeq \frac{\pi^2}{6} + \frac{\pi}{\epsilon}
\left(\ln\frac{\pi}{\epsilon} - 1 \right) + {\cal O}\left(\epsilon^{-2}\right)\,,
\,\,\, \epsilon\gg 1\,.
\label{a.20}\end{eqnarray}
With Eq.~(\ref{a.20}) the large-$\epsilon$ expansion of the constant-field rate~(\ref{a.6}) reads finally,
\begin{eqnarray}
w^{\rm cf} (\epsilon) =\frac{c}{4\pi^{3}\blambda^4_{e}}\,{\epsilon}^{2}\,{\rm Li}_{2}(e^{- \pi/\epsilon})
\simeq\frac{c}{4\pi^{3}\blambda^4_{e}}\,\epsilon^{2}
\left[\frac{\pi^2}{6} + \frac{\pi}{\epsilon}\left(\ln\frac{\pi}{\epsilon}
- 1 \right) + {\cal O}\left(\epsilon^{-2}\right)\right]\,,\,\,\, \epsilon\gg 1\,.
\label{a.21}\end{eqnarray}
The large-$\epsilon$ behavior of the rate~(\ref{a.12}) we find from the expansion~(\ref{4.31}) in the limit
$\beta\rightarrow\infty$ with fixed $\epsilon\gg 1$. In this limit, substituting $\Lambda\rightarrow \pi/\epsilon\ll 1$
into Eq.~(\ref{4.31}) yields
\begin{eqnarray}
w^{\rm lcf} (\epsilon) &\simeq&
\frac{ c}{3\pi^3 \blambda_e^4}\,\epsilon^2\Bigg\{\frac{\pi^2}{6} + \left(\frac{\pi}{\epsilon}\right)
\Bigg[\frac{3}{2}\ln\frac{\pi}{\epsilon} - 0.5794415\dots\Bigg]\Bigg.\nonumber\\
&+& \Bigg.\left(\frac{\pi}{\epsilon}\right)^{2}\Bigg[\frac{3}{16}\left(\ln\frac{\pi}{\epsilon} + \gamma\right)
- 0.5411802\dots\Bigg]
+ {\cal O}\Bigg(\left(\frac{\pi}{\epsilon}\right)^{3}\Bigg)\Bigg\},\, \epsilon\gg 1,
\label{a.22}\end{eqnarray}
where $\gamma\simeq 0.5772157\dots$ is the Euler constant~\cite{abr}.
The large-$\epsilon$ asymptotic~(\ref{a.22}) is compared to the corresponding asymptotic~(\ref{a.21})
in Fig.~6b. We observe that the Schwinger formula~(\ref{a.6}) is overestimated by the locally constant-field
expression~(\ref{a.12}) as long as $\epsilon\gg 1$, i.e. in the strong-field limit.
\begin{figure}[h]
\centering
\hspace*{-2.2cm}
\includegraphics[width=200mm,angle=0]{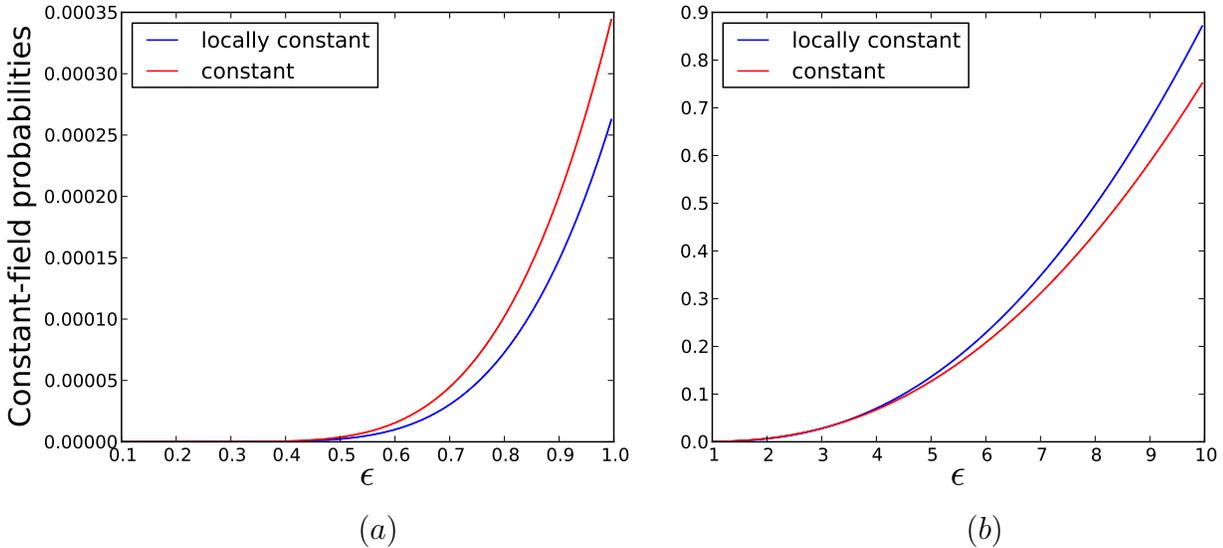}
\begin{picture}(0,10)
\put(-110,-2){$(a)$}
\put(120,-2){$(b)$}
\end{picture}
\caption{\label{Fig6} The dimensionless probability $w\cdot\left(\blambda^{4}_{e}/c\right)$
to produce an $e^{+}e^{-}$ pair within one Compton space-time volume $\blambda_{e}^{4}/c\approx 10^{-58}\,{\rm m}^{3} {\rm s}$
as a function of the constant electric field $\epsilon = E_{0}/E_{c}$. The weak-field regime $\epsilon\ll 1$
is shown in (a), the strong-field $\epsilon\gg 1$ in (b). The solid red and blue lines refer to the Schwinger
formula~(\ref{a.6}) and the expression~(\ref{a.12}), respectively.}
\end{figure}

\end{document}